\documentclass[12pt]{article}
\usepackage{jheppub_modified_green}
\usepackage{graphicx}
\usepackage{epsfig}
\usepackage{amsfonts}
\usepackage{amssymb}
\usepackage{amsmath}
\usepackage{amsthm}
\usepackage{subfig}
\usepackage{bm}
\usepackage{hyperref}
\usepackage{mathrsfs}
\usepackage{bbm}
\usepackage{cancel}
\usepackage{color}
\usepackage{accents}

\usepackage[T1]{fontenc} 

\newcommand{\Res}{ \text{Res}}
\newcommand{\updown}[2]{{\makebox [0 pt ]{${\scriptscriptstyle{#1\atop #2}}$}}}

\title{\boldmath A Combinatoric Shortcut to Evaluate CHY-forms}

\author[a,b]{Gang Chen}
\author[c]{Yeuk-Kwan E. Cheung}
\author[c]{Tianheng Wang}
\author[d]{Feng Xu}

\affiliation[a]{Department of Physics and Astronomy, Uppsala University, Uppsala, Sweden}
\affiliation[b]{Department of Physics, Zhejiang Normal University, Jinhua, Zhejiang Province, China}
\affiliation[c]{Department of Physics, Nanjing University, Nanjing, Jiangsu Province, China}
\affiliation[d]{Weavi Corporation Limited, Nanjing, Jiangsu Province, China}

\emailAdd{gang.chern@gmail.com}
\emailAdd{cheung@nju.edu.cn}
\emailAdd{tianhengwang@outlook.com}
\emailAdd{schyfeng@gmail.com}

\abstract{In \cite{Chen:2016fgi} we proposed a differential operator 
for the evaluation of the multi-dimensional residues  on isolated (zero-dimensional) poles. 
 In this paper we discuss some new insight  on  evaluating the (generalized) Cachazo-He-Yuan (CHY) forms of the scattering amplitudes using this differential operator.  
We introduce a tableau representation for the coefficients 
appearing in the proposed differential operator. 
Combining the tableaux with  the polynomial forms of the scattering equations,  the evaluation of the generalized CHY form  
 becomes a simple combinatoric  problem. 
It is thus possible to obtain the coefficients arising in the differential 
operator in a straightforward  way. We present the procedure for  
a complete solution of the $n$-gon amplitudes 
at one-loop level in a generalized CHY form.  
We also apply our method to fully evaluate  the one-loop five-point amplitude 
in the maximally  supersymmetric  Yang-Mills theory; 
 the final result is identical to the one obtained by  Q-Cut.
}

\begin{document} 
\maketitle
\flushbottom

\section{Introduction}
\label{sec:intro}
Scattering equations are derived at tree level for the high-energy behavior of string theory in~\cite{Gross:1989ge,Caputa:2011zk} and have drawn attention from theoretical physicists in diversified contexts~\cite{Witten:2004cp,Cachazo:2012uq}. 
Incorporating scattering equations, Cachazo, He and Yuan~\cite{Cachazo:2013hca, Cachazo:2013iea, Cachazo:2014nsa}  propose a closed formula for arbitrary n-point tree amplitudes in a variety of massless quantum field theories.
This form is proven in~\cite{Dolan:2013isa} for Yang-Mills theory in arbitrary dimensions and a polynomial form of the scattering equations is obtained by the same group in~\cite{Dolan:2014ega}. 

The scattering equations are generalized to loop levels in a number of contexts, such as open string theory, pure spinor formalism of superstring, and ambitwistor string theory~\cite{Berkovits:2013xba, Mason:2013sva, Geyer:2014fka}. The CHY expressions are subsequently extended to one and two loops for the bi-adjoint scalar theory, gauge theory and gravity in~\cite{Casali:2015vta,Geyer:2015bja, Geyer:2015jch, Geyer:2016wjx}.
In addition to the ambitwitor approach, the one-loop generalized CHY forms are obtained from tree-level ones in one higher spatial dimension in~\cite{Cachazo:2015aol,He:2015yua} for scalar and gauge theories, and higher-loop CHY forms for scalar theory are constructed in~\cite{Feng:2016nrf}.

Besides constructing the generalized CHY forms, it is also a challenge to evaluate such multi-dimensional CHY integrals.
Indirect methods, such as ``building block method"~\cite{Cachazo:2015nwa,Gomez:2016bmv,Cardona:2016bpi, Gomez:2016cqb} and  ``Integration rules"~\cite{Baadsgaard:2015voa,Lam:2015sqb,  Lam:2016tlk, Baadsgaard:2015hia,Mafra:2016ltu, Huang:2016zzb,Bjerrum-Bohr:2016juj, Cardona:2016gon}, evaluate the generalized CHY integrals reductively. 
Direct approaches are also explored. At tree level, the n-point CHY form is evaluated for gauge and gravity theories in special kinematics in~\cite{Kalousios:2013eca} and the scattering equations are solved in four dimensions up to six points in~\cite{Weinzierl:2014vwa, Lam:2014tga, Kalousios:2015fya}. Elimination theory is exploited in solving the scattering equations up to seven points in~\cite{Cardona:2015eba, Cardona:2015ouc} and a general prescription based on elimination theory is proposed in~\cite{Dolan:2015iln}. A direct evaluation of the CHY form for the MHV tree amplitude is given in~\cite{Du:2016blz}.
Algebraic geometry based methods--the companion matrix method~\cite{Huang:2015yka},  Bezoutian matrix method~\cite{Sogaard:2015dba},  and polynomial reduction techniques \cite{Bosma:2016ttj,Zlotnikov:2016wtk}--are employed to evaluate the CHY expressions, without solving the underlying 
scattering equations.

In our previous work \cite{Chen:2016fgi} we proposed a conjecture that enables us to compute multidimensional residues on isolated poles (0-dimension) by a differential operator. Here we briefly recall that conjecture.

Suppose $f_1$, $f_2$, ..., $f_k$ are homogeneous polynomials in complex variables $z_1$, $z_2$, ... $z_k$ of degrees $d_1$, $d_2$, ..., $d_k$ respectively. And we assume that the common zeros of $f_1$, $f_2$, ..., $f_k$ consists of a single isolated point $p$. Let $\mathcal{R}(z_i)$ be a holomorphic function in a neighborhood of $p$. Then the conjecture states that the residue of $\mathcal{R}$ at $p$ can be computed by a differential operator $\mathbb{D}$ as follows,
\begin{align}
\Res_{\{(f_1),\cdots,(f_k)\},p}[\mathcal{R}]
\equiv\oint \frac{dz_1\wedge \cdots \wedge dz_k}{f_1\cdots f_k} \mathcal R 
=\left. \mathbb{D}[\mathcal R]\right|_{z_i\rightarrow 0}, 
\end{align}
where $\mathbb D$ is of the form
\begin{align}
\mathbb{D} 
=\sum_{\{r_i\}}a_{r_1, r_2 ,\cdots, r_{k}}
 (\partial_{1})^{r_{1}}(\partial_{2})^{r_{2}}\cdots 
  (\partial_{k})^{r_{k}}\,.\label{defD}
\end{align}
Here coefficients $a_{r_1,r_2,\dots,r_k}$ are $z$-independent constants and $\partial_i=\frac{\partial}{\partial z_i}$, $i=1,\dots,k$. The sum is done over all solutions of the equation $\sum_{i=1}^{k} r_i= \sum_{h=1}^k d_h -k$. Moreover, it is conjectured that $\mathbb D$ is uniquely determined by two conditions respectively from 1) the local duality theorem \cite{hartshorne2013algebraic, griffiths2014principles} and 2) the intersection number of the divisors $D_i=(f_i)$.

This conjecture is verified numerically to be widely applicable for computing both degenerate and non-degenerate multi-dimensional residues, as long as the poles are  isolated. 
In the application to evaluating the generalized CHY integrals, the linear equations for $a_{r_1,\cdots, r_k}$ arising from the local duality theorem and the intersection number requirement are the structures of interest. 
In this paper, we study these  structures by introducing a tableau representation for each $a_{r_1,\cdots, r_k}$.
Using the tableau representation, we explain how to obtain the differential operator {efficiently}, especially for the polynomial scattering equations, which makes the evaluation of the generalized CHY forms a simple combinatoric problem.
{In this way, the evaluation of CHY integrals in their \emph{prepared forms} (which we define later in this paper) is easily achieved for any number of external lines at one loop.}

This leads straightforwardly to a full evaluation of the $n$-gon one-loop integrand in the generalized CHY form. 
Furthermore, many a term in the integrands of super-Yang-Mills 
expressed in the generalized CHY form can also be re-casted into 
the prepared form, which are then calculated effortlessly.
We apply our method on the one-loop five-point 
amplitude in SYM and find that the long {expression} in the Pfaffian can be transformed into the prepared form by cross-ratio identities~\cite{Cardona:2016gon}. 
Upon evaluating the prepared forms, the final result is identical to the ones obtained in~\cite{Carrasco:2011mn} by Q-Cuts~\cite{Baadsgaard:2015twa}.

\section{Prescription for determining differential operators}
\label{sec:prescription}
In this section we introduce our new method to determine the differential operators that is more efficient than the approach taken in \cite{Chen:2016fgi}. We begin with a warmup example for the four point one loop SYM integrand. After that we present the notations used in obtaining the coefficients in the differential operators. Finally we use this method to get the complete solutions of the coefficients for a class of CHY integrands. 
\subsection{Toy model: one-loop four-point SYM integrand}
The one-loop scattering amplitude for four external gluons in $\mathcal N=4$ SYM has been revisited in various contexts in literature~\cite{Bern:1994zx,Carrasco:2011mn}. The generalized CHY integral for its loop integrand is given in~\cite{Geyer:2015bja, Geyer:2015jch}. A detailed analysis of the evaluation of the generalized CHY integral has already been presented in \cite{Chen:2016fgi}. In this section we reconsider this example as a toy model to illustrate the main ideas of the method proposed in this paper. Details of and more general discussions on the method are to appear in later sections.

Firstly let us recall the settings. The four-point generalized CHY \cite{Geyer:2015bja, Geyer:2015jch} form of the integrand reads, 
\begin{eqnarray}\label{eq:MasonI40}
\mathcal I_4 ={\langle 12\rangle^4\over \langle 12\rangle\langle 23\rangle\langle 34\rangle\langle 45\rangle \langle 51\rangle}\oint {d\sigma_1\cdots d\sigma_{3} \over  h_1\cdots h_{3}} ~PT_4 \prod_{i<j}^4 (\sigma_i-\sigma_j) \,,
\end{eqnarray}
where $PT_4$ is the well-known Parke-Taylor factor that reads
\begin{equation}
    PT_4=\sum_{\rho\in \mathcal{S}_4}{1\over \sigma_{\rho(1)} (\sigma_{\rho(1)}-\sigma_{\rho(2)})(\sigma_{\rho(2)}-\sigma_{\rho(3)})(\sigma_{\rho(3)}-\sigma_{\rho(4)})}\,.
\end{equation}
where $\mathcal{S}_4$ is the cyclic permutation group of four objects.
For illustrative purpose, here we only consider the first term in $PT_4$ which leads to the following integral
\begin{eqnarray}\label{eq:MasonI41}
\mathcal I_4^1 ={\langle 12\rangle^4\over \langle 12\rangle\langle 23\rangle\langle 34\rangle\langle 45\rangle \langle 51\rangle}\oint {d\sigma_1\wedge \cdots\wedge  d\sigma_{3} \over  h_1\cdots h_{3}} ~{(\sigma_{3})(\sigma_{2}-\sigma_{4})\over \sigma_{1}}  \,,
\end{eqnarray}
where we fix the gauge $\sigma_4=1$ and the polynomials $h_i$'s are
\begin{eqnarray}
h_1&=&\sum_{i=1}^4 l_i \sigma_i\nonumber, ~~~h_2=(-)\sum_{i_1<i_2}^4 \sigma_{i_1}\sigma_{i_2} l_{i_1i_2}, ~~~~h_3=\sum_{i_1<i_2<i_3}^4 \sigma_{i_1\cdots i_3} l_{i_1\cdots i_3}
\label{eq:polynomialScatEq}
\end{eqnarray}
and we have used the following notations,
\begin{equation*}
\sigma_{i_1\cdots i_m}\equiv \prod_{r=1}^m\sigma_{r}\,,\quad l_i\equiv l\cdot k_i\,, \quad l_{i_1\cdots i_m}\equiv \left(l\cdot k_{i_1\cdots i_m}{ -\frac{1}{2}} k_{i_1\cdots i_m}^2\right), \quad k_{i_1\cdots i_m}\equiv\sum_{r=1}^m k_{i_r}\,. 
\end{equation*}
As shown in \cite{Chen:2016fgi}, using the global residue theorem, the integral can be rewritten as the following
\begin{eqnarray}\label{eq:MasonI42}
\mathcal I_4^1 ={-\langle 12\rangle^4\over \langle 12\rangle\langle 23\rangle\langle 34\rangle\langle 45\rangle \langle 51\rangle}\oint {d\sigma_1\wedge \cdots\wedge  d\sigma_{3}\wedge d\sigma'_4 \over  \tilde h_1\cdots \tilde h_{3}  \sigma_{1}} ~{\sigma_{3}(\sigma_{2}-\sigma'_{4})\over \sigma'_4-1}  \,,
\end{eqnarray}
where $\tilde h_i$ is obtained from $h_i$ with the replacement $\sigma_4\rightarrow\sigma'_4$. From now on in this section we will simply write $\sigma_4$ for $\sigma_4'$ for notational convenience.

The main idea of the method presented in \cite{Chen:2016fgi} is to replace the contour integration
$$Res_{(\tilde h_1),(\tilde h_2),(\tilde h_3), (\sigma_1)}=\oint {d\sigma_1\wedge \cdots\wedge  d\sigma_{3}\wedge d\sigma'_4 \over  \tilde h_1\cdots \tilde h_{3}  \sigma_{1}}$$
with a differential operator
$\mathbb{D}$, which should be of the third-order in this particular case and can be written as
\begin{eqnarray} 
\mathbb D = \sum_{\updown{0\leq r_i\leq 3\,,} {\;r_1+r_2+r_3+r_4=3}} a_{r_1, r_2, r_3, r_4} \left(\frac{\partial}{\partial\sigma_1}\right)^{r_1} \left(\frac{\partial}{\partial\sigma_2}\right)^{r_2} \left(\frac{\partial}{\partial\sigma_3}\right)^{r_3} \left(\frac{\partial}{\partial\sigma_4}\right)^{r_4} \,,
\end{eqnarray}
where the coefficients $a$'s are to be fixed by the local duality theorem and the intersection number equation. One of the main points of the current paper is to provide an efficient method to determine those coefficients without solving the equations by brute force.

\begin{figure}[htbp]
  \centering
  \includegraphics[width=0.75\textwidth]{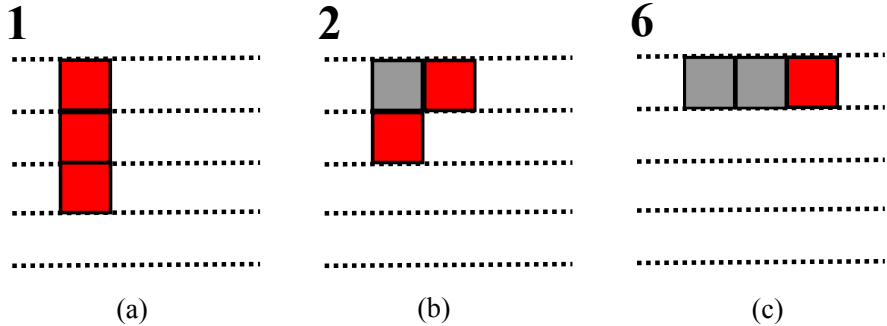}
  \caption{Tableaux for the four-point scattering equations. The red tiles in the tableaux are allowed to perform a permutation { among the rows. The meaning of such permutations are to be discussed shortly.}.}\label{fig:FourPointAll}
\end{figure}

Now we briefly describe how the shortcut method works out in this particular example.
There are 20 coefficients $a_{r_1, r_2, r_3, r_4}$ to be determined. Before turning to the equations from local duality and intersection number, we firstly classify the coefficients as follows. For each coefficient $a$ we define its \emph{rank} to be $R(a_{r_1,r_2,r_3,r_4})=r_1!r_2!r_3!r_4!$.  Furthermore, we associate a tableau with each $a$. All tableaux (up to subscript permutation) for our example are shown in Fig \ref{fig:FourPointAll}. The tableau in Fig \ref{fig:FourPointAll}(a) (including the permutations of the red tiles) denotes $a_{0,1,1,1}$ and all those with permuted subscripts. In Fig  \ref{fig:FourPointAll}(b), the corresponding variables are  $a_{0,0,1,2}$ and all those with permuted subsripts. In Fig  \ref{fig:FourPointAll}(c), the corresponding variables are  $a_{0,0,0,3}$ and all the permutations $\rho$ of the index $\{0,0,0,3\}$.

With the definition of the rank of coefficients and the tableaux representation, solving the equations of the $a$'s can be carried out in a very specific way that is much more efficient than treating it as a general equation solving problem. Recall that the local duality theorem yields,
\begin{eqnarray}
\mathbb D \left( \sigma_i \sigma_j \sigma_1 \right) = \mathbb D \left(\sigma_i \sigma_j \tilde h_1\right) = \mathbb D \left(\sigma_i \tilde h_2 \right) = \mathbb D \; \tilde h_3 =0\,\quad\quad 0\leq i,j \leq 3\,.
\end{eqnarray}
We will see that the equation $\mathbb D \left( \sigma_i \sigma_j \sigma_1 \right)=0$ simply leads to the vanishing of all $a_{r_1,\_,\_,\_}$ with $r_1\geq 1$. So this means, for a non-zero $a$, the number of tiles in the first row in the corresponding tableau is zero. On the other hand, the equation $\mathbb D \; \tilde h_3 =0$ means the vanishing of $a$'s of rank 1.  Thus $a$'s corresponding to the tableau in Fig. \ref{fig:FourPointAll} (a) are all vanishing. The equation $\mathbb D \left(\sigma_i \sigma_i \tilde h_1\right)$  will give us a relation between $a$'s of rank 6 and $a$'s of rank 2, e.g
\begin{eqnarray}
\label{eq:a003}
a_{0,0,0,3} = {-2!\over 3! \ell_4}(\ell_1 a_{1,0,0,2}+\ell_2 a_{0,1,0,2}+\ell_3 a_{0,0,1,2})\,.
 \end{eqnarray}
 This equation can actually be read off directly from the tableau in Fig. \ref{fig:FourPointAll} (c). By permuting the red tiles of $a_{0,0,0,3}$'s tableau among the rows, we could get the tableaux corresponding to $a_{1,0,0,2}$, $a_{0,1,0,2}$ and $a_{0,0,1,2}$ respectively. In terms of the coefficients, this exactly corresponds to (\ref{eq:a003}). Namely, $a_{0,0,0,3}$ multiplied by its rank and $l_4$ equals the minus sum of $a_{1,0,0,2}$, $a_{0,1,0,2}$ and $a_{0,0,1,2}$ multiplied by their ranks and $l_i$, where $i$ indicates the row index of the red tile of the corresponding tableau.
 
\begin{figure}[htbp]
  \centering
  \includegraphics[width=0.45\textwidth]{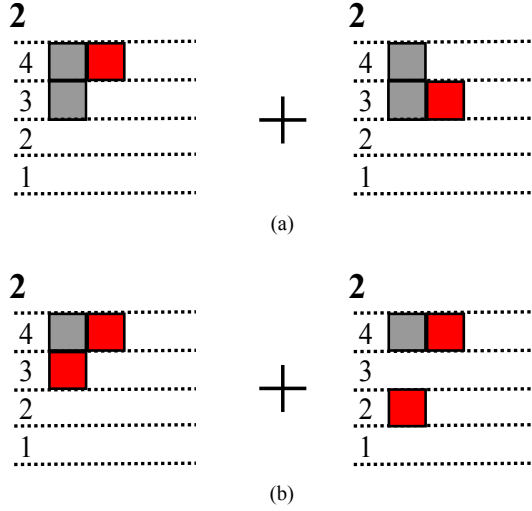}
  \caption{Tableaux indicating the permutations of the red tiles among the rows.}\label{fig:FourPointMove}
\end{figure}

The remaining unknown $a$'s are those corresponding to the tableau Fig. \ref{fig:FourPointAll} (b) with rank 2.  They are all the variables present in the intersection number equation
 \begin{eqnarray}
 \label{eq:IntN}
&&2  a_{0,0,1,2}  \ell_{4} \ell_{3,4}\ell_{2,3,4}-2  a_{0,0,2,1}  \ell_{3} \ell_{3,4}\ell_{2,3,4}+2  a_{0,2,0,1}  \ell_{2} \ell_{2,4}\ell_{2,3,4}\nonumber\\
&-&2  a_{0,1,0,2}  \ell_{4} \ell_{2,4}\ell_{2,3,4}+2  a_{0,1,2,0}  \ell_{3} \ell_{2,3}\ell_{2,3,4}-2  a_{0,2,1,0}  \ell_{2} \ell_{2,3}\ell_{2,3,4}=6.
\end{eqnarray}
Other equations involving variables of rank two are $\mathbb D \left(\sigma_i \sigma_j \tilde h_1\right)$ with $i\neq j$ and  $\mathbb D \left(\sigma_i \tilde h_2\right)$.  For example $\mathbb D \left(\sigma_3 \sigma_4\tilde h_1\right)=0$ is
\begin{eqnarray}
\label{eq:a011a}
2 \ell_4 a_{0,0,1,2} + 2\ell_3 a_{0,0,2,1}=0,
 \end{eqnarray}
 where we have used the fact that $a_{0,1,1,1}$ and $a_{1,0,1,1}$ are zero.
 This equation is represented by Fig. \ref{fig:FourPointMove}(a) in the same way as that for Eq. (\ref{eq:a003}). This equation tells us that the first two terms in the Eq. (\ref{eq:IntN}) are the same. The equation $\mathbb D \left(\sigma_4 \tilde h_2\right)=0$ is, after ignoring the zero variables,
 \begin{eqnarray}
\label{eq:a011b}
2 \ell_{3,4} a_{0,0,1,2} + 2\ell_{2,3} a_{0,1,0,2}=0.
 \end{eqnarray}
Similarly this equation can also be read off from the tableau. For that purpose, we paint red the two tiles as shown in Fig.  \ref{fig:FourPointMove}(b) and permute the red tiles among the four rows. To get Eq. (\ref{eq:a011b}), we simply take the sum of the products of the corresponding $a$'s with their ranks and corresponding $l_{i,j}$'s, where $i,j$ are the row indices of the red tiles of respective tableaux.  Accordingly, it is also seen that the first term and the forth term on the left hand side of Eq. (\ref{eq:IntN}) are equal.
By considering all equations of $\mathbb D \left(\sigma_i \sigma_j \tilde h_1\right)=0$ with $i\neq j$ and  $\mathbb D \left(\sigma_i \tilde h_2\right)=0$ in a similary way,  we can see that all the terms in the intersection number equation are actually equal. Hence the solutions of those $a$'s appearing in (\ref{eq:IntN}) are just the inverses of their respective coefficients. Up to this point, we thus have already solved all the $a$'s and found the exact form of the operator $\mathbb D$. Now we could apply the $\mathbb D$ operator to evaluate the generalized CHY integrand,
\begin{eqnarray}\label{eq:MasonI45}
\mathcal I_4^1 &=&{-\langle 12\rangle^4\over \langle 12\rangle\langle 23\rangle\langle 34\rangle\langle 45\rangle \langle 51\rangle}\oint {d\sigma_1\wedge \cdots\wedge  d\sigma_{3}\wedge d\sigma_4 \over  \tilde h_1\cdots \tilde h_{3}  \sigma_{1}} ~{\sigma_{3}(\sigma_{2}-\sigma_{4})\over \sigma_4-1}  \nonumber\\
&=&{-\langle 12\rangle^4\over \langle 12\rangle\langle 23\rangle\langle 34\rangle\langle 45\rangle \langle 51\rangle}2a_{0,0,1,2}={-\langle 12\rangle^4\over \langle 12\rangle\langle 23\rangle\langle 34\rangle\langle 45\rangle \langle 51\rangle}\ell_{4} \ell_{3,4}\ell_{2,3,4}\,.
\end{eqnarray}

\subsection{Preliminary: notations \& tableaux}
In the previous discussion, the one-loop SYM integrand for four external particles is massaged and turned into a sum of several terms, all of which are of the following form,
\begin{align}\label{eq:prepared}
\mathcal I_n^{l=1} = \oint\limits_{h_1 = \cdots h_{n-1}= h'_n = 0}\frac{d\sigma_1 \wedge \cdots \wedge d\sigma_{n-1} \wedge d\sigma'_n}{h_1 \cdots h_{n-1} h'_n}\, \mathcal H(\sigma_1,\dots, \sigma_{n-1},\sigma'_n)\,,
\end{align}
where $\sigma_j~(j=1,\cdots, n-1)$ are the $(n-1)$ variables that need to be integrated out (with the gauge choice $\sigma_n=1$) and $\sigma'_n$ is the auxiliary parameter that comes into play in the homogenization of the scattering equations {(for details, see Section 4.2 and 4.3 of \cite{Chen:2016fgi})}. $h_j~(j=1,\cdots , n-1)$ denote the $(n-1)$ polynomial scattering equations (\ref{eq:ScatEq}) that are necessary to capture the behavior of an $n$-point scattering and $h'_n = \sigma_i~(i=1,\cdots, n-1)$ or $h'_n = \sigma'_n$. In this form, the divisors are generated either by one of the polynomial scattering equations or by the simplest monomial possible. As reviewed in the introduction, such contour integrals are associated with a differential operator with a certain number of coefficients fixed by the local duality theorem and the intersection number. These constraints are all linear and especially the local duality theorem gives rise a sparse constraint matrix, which in general can be solved by various sparse matrix methods. In the case of an integral in the form of (\ref{eq:prepared}), thanks to the beautiful mathematical structures of scattering equations, these constraints can indeed be solved analytically. In this section we present the method that reduces considerably the size of the constraint matrix and obtains quickly the analytic solutions of the coefficients. 
For this reason, we would like to call an expression of form (\ref{eq:prepared}) \textit{prepared}. Moreover, as we will observe in more examples, the CHY-like representations for amplitudes/integrands can often be reduced to the prepared form or similar ones with a slightly modified $h'_n$, even though the original seem much more complicated. For an expression with a modified $h'_n$ which still remains a monomial, we expect our algorithm can be generalized.

\begin{figure}[htbp]
  \centering
  \includegraphics[width=0.45\textwidth]{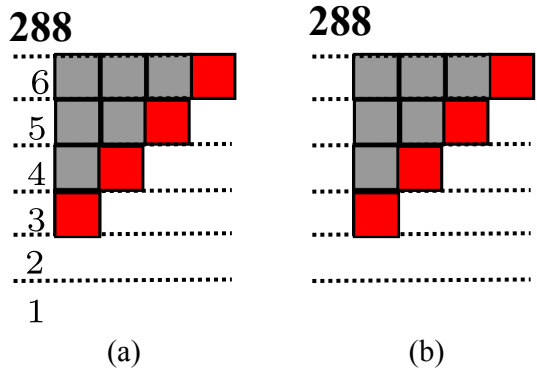}
  \caption{Box Diagram.}\label{fig:specified}
\end{figure}

As shown in the discussion of the four-point one-loop integrand, to demonstrate our  method in a more intuitive way, a graphical representation of the coefficients in the operator comes in handy. Let us take a coefficient $a_{0,1,2,3,4}$ as an example and address a few details of its corresponding tableau depicted in Fig.\ref{fig:specified} (a). 
The $i$-th row of the tableau corresponds to the $i$-th index of the coefficient while the number of the tiles in the $i$-th row is equal to the value of the $i$-th index. The total number of tiles is the order of the corresponding differential operator. As mentioned above, we define the \textit{rank} of as coefficient $a_{r_1, r_2, \cdots, r_n}$ as the following integer number,
\begin{align}\label{eq:DefRank}
R(a_{r_1, r_2, \cdots, r_n}) = r_1!\, r_2 !\cdots r_n!\,,
\end{align}
and this number is written on top of the corresponding tableau. 
Sometimes we drop the row labels in the tableau as shown in Figure.\ref{fig:specified}(b), and this tableau represents the class of $a$'s whose indices are related by permutations. For instance, the tableau in Figure.\ref{fig:specified}(b) is corresponding to the coefficients $ \{  a_{\rho(0,0,1,2,3) }  \}_{\rho\in S_5}$ where $S_5$ is the permutation group of five items and $\rho(0,0,1,2,3)$ means a permuation of the five digits $(0,0,1,2,3)$.

\begin{figure}[htbp]
  \centering
  \includegraphics[width=0.65\textwidth]{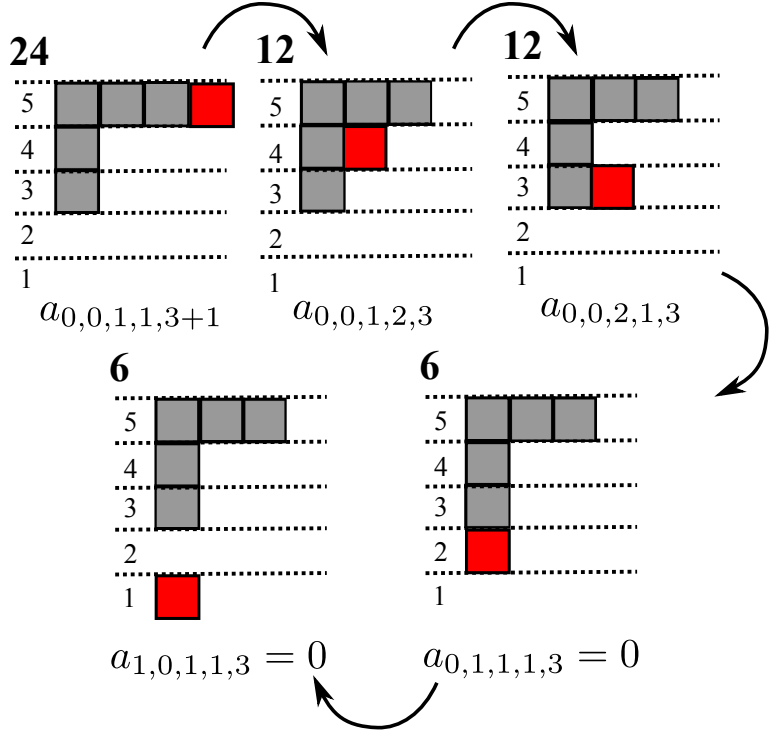}
  \caption{Example of a tile move.}\label{fig:a411Move}
\end{figure}

As shown in Figure.\ref{fig:specified}, we have painted some tiles red in the tableaux. For a given unpainted tableau, we associate with it a class of colored tableaux which are obtained in the following way: in each row only the rightmost tile can be painted red, and we can choose to paint it or not. So altogether we have $2^J$ colored tableaux for each given unpainted one, where $J$ is the number of nonempty rows in the unpainted tableau. From each colored tableau, we can get several new tableaux (without considering their coloring) by permuting the red tiles among the rows. For example,  in Figure.\ref{fig:a411Move} are depicted the new tableaux resulted from the tile moving of the first colored tableau. In this example, we can obtain $a_{0,0,1,2,3}$, $a_{0,0,2,1,3}$, $a_{0,1,1,1,3}$ and $a_{1,0,1,1,3}$ from $a_{0,0,1,1,4}$. 

So far we have played with the tableaux. Now we demonstrate how the tile moving can acquire actual meanings from the local duality theorem. Recall that the polynomial scattering equation of degree $m$ for $n$ particles takes the following form (throughout this paper we adopt the gauge choice $\sigma_n =1$ and all the polynomial scattering equations, if not specified otherwise, are homogenized with an auxiliary parameter $\sigma'_n$; for notational compactness, we drop the prime and simply denote the auxiliary as $\sigma_n$ from now on),
\begin{align}
h_m = \sum_{1\leqslant i_1 < i_2 <\cdots < i_m\leqslant n} \ell_{i_1 i_2 \cdots i_m} \sigma_{i_1} \sigma_{i_2} \cdots \sigma_{i_m}\,.
\end{align}
The order of the differential operator $\mathbb D$ associated with the prepared form for $n$ points is 
\begin{align*}
M = 1+ 2 + \cdots (n-2) = {(n-1) (n-2) \over 2} \,.
\end{align*}
Since the differential operator $\mathbb D$ acting on a monomial in $\sigma_i$'s does nothing but picking out the exponents, the local duality theorem yields,
\begin{eqnarray}\label{eq:LocalGen}
&& \mathbb D \left( \sigma^{r_1}_1 \sigma^{r_2}_2 \cdots \sigma^{r_n}_{n} \, h_m\right) \nonumber\\
&=& \sum_{i_1 < i_2 <\cdots i_m} \ell_{i_1 i_2 \cdots i_m} \underbrace{\left(r_1! \cdots (r_{i_1}+1)! \cdots (r_{i_m}+1)!\cdots r_n!\right)}_{R(a_{r_1, \cdots, (r_{i_1}+1),\cdots, (r_{i_m}+1),\cdots, r_n })} a_{r_1, \cdots, (r_{i_1}+1),\cdots, (r_{i_m}+1),\cdots, r_n }=0 \,,\nonumber\\
\end{eqnarray}
where $r_j$'s are non-negative integers satisfying the Frobenius equation $r_1+ r_2 + \cdots r_n = M-m$. Here we see that the rank of $a_{r_1,\cdots, r_n}$ is merely the factorial product that arises when taking derivatives multiple times. There are $C_n^m$ coefficients $a_{r_1,\cdots, r_n}$'s on the right-hand side above and their corresponding tableaux all have $r_j$ black tiles in the row $j$ and $m$ red tiles that sit in the $m$ different rows $\{i_1, i_2, \cdots, i_m \}$. The tableaux corresponding to these $a$'s are exactly those related by permuting the red tiles as mentioned above!
In other words, given a tableau that has $r_j$ black tiles in the $j$-th row and the rows $\{i_1,\cdots, i_m\}$ have one red tile each, there is a unique equation following from the local duality theorem that relates this tableau and those obtained by permuting the red tiles. Furthermore, scanning over the monomials of the correct degrees to construct the local duality equations is equivalent to exhausting all the legitimate colorings and writing down the corresponding equations.

As mentioned above, the equations arising from the local duality theorem are not  independent from each other and the number of equations is much larger than the number of coefficients. With the correspondence between the colorings and the local duality constraints established, we are ready to present a method that systematically selects just enough independent local duality equations to fix the coefficients completely. 

\subsection{Coefficient generating algorithm}
In this section, we spell out our method that analytically solves the local duality and the intersection number constraints associated with a prepared form and display the general solutions of the coefficients in the corresponding differential operator.

Although there are several possible colored tableaux associated with a given uncolored tableau, there is a preferred one that leads (by shuffling the red tiles among the rows) to new tableaux whose ranks are either lower than or equal to that of the original. This particular colored tableau is constructed as follows. For convenience let us assume $a_{r_1, r_2, \cdots, r_n}$ is a coefficient with indices such that $r_{i_1}\leqslant r_{i_2}\leqslant\cdots\leqslant r_{i_n}$. In the prefered colored tableau, we paint red the last tile in row $i_n$ (the top row). Now for all subsequent rows, if the row $i_p$ has one red tile and $r_{i_p}-r_{i_{p-1}}\leqslant 1,~r_{i_{p-1}}>0$, we should also paint red the last tile of the row $r_{i_{p-1}}$; otherwise all the tiles in the row $r_{i_{p-1}}$ remain black and the coloring process stops here. 

There are two types of the coefficients $a_{r_1, r_2, \cdots, r_n}$: let $(j_1, j_2, \cdots j_n)$ be a particular permutation of the indices such that $r_{j_1} \leqslant r_{j_2} \leqslant \cdots \leqslant r_{j_n}$; if $r_{j+1} - r_{j} \leqslant 1$ for $j=1,\cdots n-1$, $a_{r_1 r_2 \cdots r_n}$ is called \textit{elementary}; otherwise it is called non-elementary. It is easy to observe that, for an elementary coefficient, the aforementioned preferred coloring relates it with those of lower or equal ranks. A non-elementary coefficient gets turned into only lower-rank ones through the preferred coloring. Moreover, for a particular prepared form for $n$ points, the corresponding tableaux are all of $M=(n-1)(n-2)/2$ tiles in total. Up to row permutation, there is a unique tableau with the highest rank among the elementary ones, that is the tableau associated with $a_{0,0,1,2,3,\dots,n-2}$ whose rank is $R_e=(n-2)! (n-1)!\cdots 1!$. 

The above observations about the prepared form actually provide us a natural solving order in which the local duality constraints can be conveniently solved. 
In the following discussion, we will show that in the operator associated with a prepared form, any coefficients with ranks lower than $R_e$ must vanish while the elementary ones with rank $R_e$ can be easily worked out, which leaves us only non-elementary ones with ranks greater than $R_e$. Those higher rank coefficients can however be obtained by an inductive method: suppose we have obtained all the coefficients -- both elementary and non-elementary -- up to some rank $R_0\geqslant R_e$ and the next possible rank is $R_1$, then a non-elementary coefficient of rank $R_1$ whose tableau is painted in the preferred way can be directly read off from the relation associated with the colored tableau, i.e. it can be represented in terms of other coefficients of lower ranks only. This procedure continues all the way up to the highest-rank coefficients and thus in this way we could determine all the coefficients analytically one after another.
%

\paragraph{Trivial coefficients}
In general, the differential operator $\mathbb D$ may be very complicated. However, it is easy to see that many coefficients in $\mathbb D$ are actually zero, i.e trivial, when $\mathbb D$ is associated with a form that is prepared (see (\ref{eq:prepared})).

Firstly, for a prepared form with $h_n = \sigma_i$, the coefficients $a_{r_1, \cdots, r_n}$ with $r_i \geqslant 1$ must vanish, because the local duality theorem for $h_n$ yields,
\begin{align}
\mathbb D (\sigma_1^{r_1} \cdots \sigma_n^{r_n} h_n) = a_{1+r_1, r_2 ,\cdots, r_n} =0\,,
\end{align}
where $s_j$ are non-negative solutions to $r_1+\cdots + r_n = M-1$.

Secondly, the coefficients $a_{r_1,\cdots, r_n}$ must vanish if  $R(a_{r_1,\cdots, r_n}) < R_e$. To see this, let us consider a prepared form with $h_n = \sigma_i$. Since $r_1+\cdots+ r_n = M$, the lowest-rank coefficients must be elementary and have $r_i\geqslant 1$ when $n\geqslant 5$ (when $n=4$ the lowest-rank coefficient has the $i$-th index being 0 and all the rest 1), which obviously vanish.  
Now we only need to show that every elementary coefficient of the form $a_{r_1,\cdots, r_{i-1},0,r_{i+1}\cdots r_n}$ (where $r_1!\cdots r_n!<R_e$) is zero, since the non-elementary ones can be written in terms of coefficients with lower ranks only. 
Let $\{ i_1, \cdots , i_{i-1}, i_{i+1},\cdots , i_n \}$ be a permutation of $\{1,\cdots, i-1, i+1,\cdots,  n\}$ such that $r_{i_1} \leqslant \cdots \leqslant r_{i_n}$ and $r_{i_{j+1}} - r_{i_j} \leqslant 1$, $r_{i+1} - r_{i-1} \leqslant 1$. 
Then we must have $r_{i_n} \leqslant n-3 $ in order to guarantee that the rank is lower than $R_e$. \footnote{If $r_{i_n} = n-2$ and all the other $r_j$'s saturate their lower bounds, we have
$$ R(a_{0, r_2\cdots r_n}) = r_{i_n}! r_{i_{n-1}}! \cdots r_{i_1}! = (n-2)! (n-3)! \cdots 1!\,.$$}
On the other hand, we must also have $r_{i_1}=1$ because $r_2+\cdots + r_n = M=\text{ord}(\mathbb D)$.\footnote{
Likewise, if $r_1=0$ and all the other $r_j$'s saturate their upper bounds, we have
$$ r_2+\cdots + r_n = r_{i_2}+\cdots + r_{i_n} = 0+1+\cdots+ (n-3) + (n-3) < \frac{(n-1)(n-2)}{2} = M\,.$$ 
Therefore we must have $r_{i_1} = 1$ and $r_{i_n} \leqslant n-3$.
}
Hence the tableau for $a_{r_1,\cdots, r_{i-1}, 0, r_{i+1},\cdots, r_n}$ has only the $i$-th row empty and all the other rows filled. The aforementioned preferred coloring leads to $(n-1)$ red tiles in total and the local duality relation represented by this colored tableau involves $a_{r_1,\cdots, r_{i_1}, 0, r_{i+1},\cdots, r_n}$ and other coefficients that all have $r_i=1$, i.e. $a_{r_1,\cdots, r_{i-1}, 0, r_{i+1},\cdots, r_n}$ is rewritten as a linear combination of several vanishing $a$'s and therefore is zero. By induction, all the non-elementary coefficients with ranks lower than $R_e$ are also zero.

\begin{figure}[htbp]
  \centering
  \includegraphics[width=0.25\textwidth]{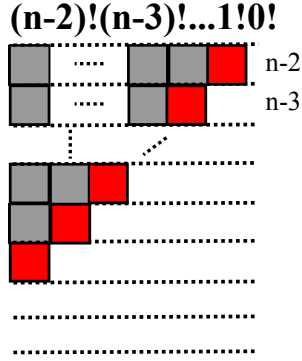}
  \caption{ The  tableaux  $a_{\{0,0,1,2,\cdots,n-3,n-2\}},$}\label{fig:IntersectA}
\end{figure}

\paragraph{Non-trivial elementary coefficients} Now we turn to the elementary coefficients of rank $R_e$. Recall that the intersection number constraint reads
\begin{align}\label{eq:intersection_number_eq}
\mathbb{D}[\det(\partial_i h_j)]=(n-1)! \,,
\end{align}
where $h_n =\sigma_i$ and the scattering equations $h_j~(j=1,\cdots n-1)$ are polynomials in $\sigma_i$'s of degree $j$ respectively. Using the explicit forms of $h$'s, we can see that only coefficients of ranks no greater than $R_e$ appear in (\ref{eq:intersection_number_eq}).  Combining with the previous discussion about trivial coefficients, we know that the only surviving coefficients in the intersection number equation are of the form
\begin{align}
a_{\rho(0),\rho(1),\dots,\rho(i-2),0,\rho(i-1),\dots,\rho(n-2)}
\end{align}
where $\rho$ is a permutation of $\{0,1,2,\dots,n-2\}$, i.e all $a$'s with the $i$-th index zero and other indices a permutation of $\{0,1,2,\dots,n-2\}$. These are elementary and their corresponding tableaux are shown in Figure.\ref{fig:IntersectA}. 

The intersection number equation (\ref{eq:intersection_number_eq}) is then
\begin{eqnarray}
\label{eq:IntNPer}
(n-2)!\cdots 2! 1! 0!&\sum_\rho& (-1)^{\text{sgn}(\rho)} \ell_{v^{(\rho)}_{n-2}} \ell_{v^{(\rho)}_{n-3,n-2}}\cdots \ell_{v^{(\rho)}_{1,2,\cdots,n-3,n-2}}\ell_{1,\cdots, i-1,i,\cdots, n}\nonumber\\
&\times&a_{\rho(0),\rho(1),\dots,\rho(i-2),0,\rho(i-1),\dots,\rho(n-2)} =(n-1)!
\end{eqnarray}
where $v^{(\rho)}_{j_1,\cdots, j_r}$ is the position set of the number set $\{j_1,\cdots, j_r\}$ among the indices of \\ $a_{\rho(0),\dots,\rho(i-2),0,\rho(i-1),\dots,\rho(n-2)}$.
A pair of such $a_{\rho(0),\rho(1),\dots,\rho(i-2),0,\rho(i-1),\dots,\rho(n-2)}$'s are directly related by the local duality theorem, as long as their indices are related by permutations. To see this, let us consider the following two permutations,
\begin{eqnarray}
\rho_1\{0,1,2,3\cdots,n-2\}&=&\{0,1,\cdots, n-3,\cdots,n-2\}\,,\\
\rho_2\{0,1,2,3\cdots,n-2\}&=&\{0,1,\cdots, n-2,\cdots,n-3\}\, .
\end{eqnarray}
The right-hand sides of the two equations above are associated with each other by swapping $(n-3)$ and $(n-2)$ and the local duality theorem associated with $h_1$ yields,
\begin{align}\label{eq:local_duality_eg}
\mathbb D \left( \sigma_1^{m_1} \cdots \sigma_{i-1}^{m_{i-1}}\sigma_{i+1}^{m_{i+1}}\cdots \sigma_n^{m_n} h_1\right) = 0\,,
\end{align}
where $m_j = \text{min}(\rho_1(j), \rho_2(j)),~(j=0,1,2,\cdots, n-2)$ and in particular $m_{v_{n-2}^{(\rho_1)}} = (n-3),~ m_{v_{n-2}^{(\rho_2)}}=(n-3)$. Only two terms survive on the left-hand side of (\ref{eq:local_duality_eg}) and this gives the pairwise relation among the non-trivial elementary coefficients as follows,
\begin{align}
\text{sgn}(\rho_1)\ell_{v^{(\rho_1)}_{n-2}} a_{\rho_1(0),\dots,\rho_1(i-2),0,\rho_1(i-1),\dots,\rho_1(n-2)}= \text{sgn}(\rho_2) \ell_{v^{(\rho_2)}_{n-2}} a_{\rho_2(0),\dots,\rho_2(i-2),0,\rho_2(i-1),\dots,\rho_2(n-2)}.
\end{align}
Now we consider another two permutations,
\begin{eqnarray}
\rho_1\{0,1,2,3\cdots,n-2\}&=&\{\cdots, n-4, \cdots, n-3, \cdots, n-2\}\,,\\
\rho_2\{0,1,2,3\cdots,n-2\}&=&\{\cdots, n-3,\cdots,n-2, \cdots, n-4\}\,,
\end{eqnarray}
where the right-hand sides are related by two permutation actions and the corresponding equation is given by the local duality theorem for $h_2$ which reads,
\begin{align}
\mathbb D \left( \sigma_1^{m_1}\cdots \sigma_{i-1}^{m_{i-1}}\sigma_{i+1}^{m_{i+1}}\cdots \sigma_n^{m_n} h_2\right) = 0\,,
\end{align}
where $m_j$ is defined the same way as above and for this case we have $m_{v^{(\rho_2)}_{n-3}} = (n-4)$, $m^{(\rho_2)}_{n-2} = (n-3)$ and $m^{(\rho_1)}_{n-2} = (n-4)$. This equation also results in a relation for a pair of non-trivial elementary coefficients as follows,
\begin{align}
&\text{sgn}(\rho_1)\ell_{v^{(\rho_1)}_{n-3,n-2}} a_{\rho_1(0),\dots,\rho_1(i-2),0,\rho_1(i-1),\dots,\rho_1(n-2)}\nonumber \\
=~&\text{sgn}(\rho_2) \ell_{v^{(\rho_2)}_{n-3,n-2}}a_{\rho_2(0),\dots,\rho_2(i-2),0,\rho_2(i-1),\dots,\rho_2(n-2)} \,.
\end{align}
By induction, we can construct the relations between any given pair of non-trivial elementary coefficients, since they all have the same indices up to permutations. Now it is straightforward to eliminate all but one coefficient in the intersection number equation and solve this constraint right away, which gives rise to the general solution of the non-trivial elementary coefficients as follows,
\begin{align}\label{eq:elementSol}
a_{\rho(0),\dots,\rho(i-2),0,\rho(i-1),\dots,\rho(n-2)} ={(-1)^{\text{sgn}(\rho)}\over (n-2)!\cdots 1! \ell_{v^{(\rho)}_{n-2}} \ell_{v^{(\rho)}_{n-3,n-2}}\cdots \ell_{v^{(\rho)}_{1,2,\cdots,n-3,n-2}}\ell_{1,\cdots,i-1, i+1\cdots,n}}\,.
\end{align}

\begin{figure}[htbp]
  \centering
  \includegraphics[width=1.0\textwidth]{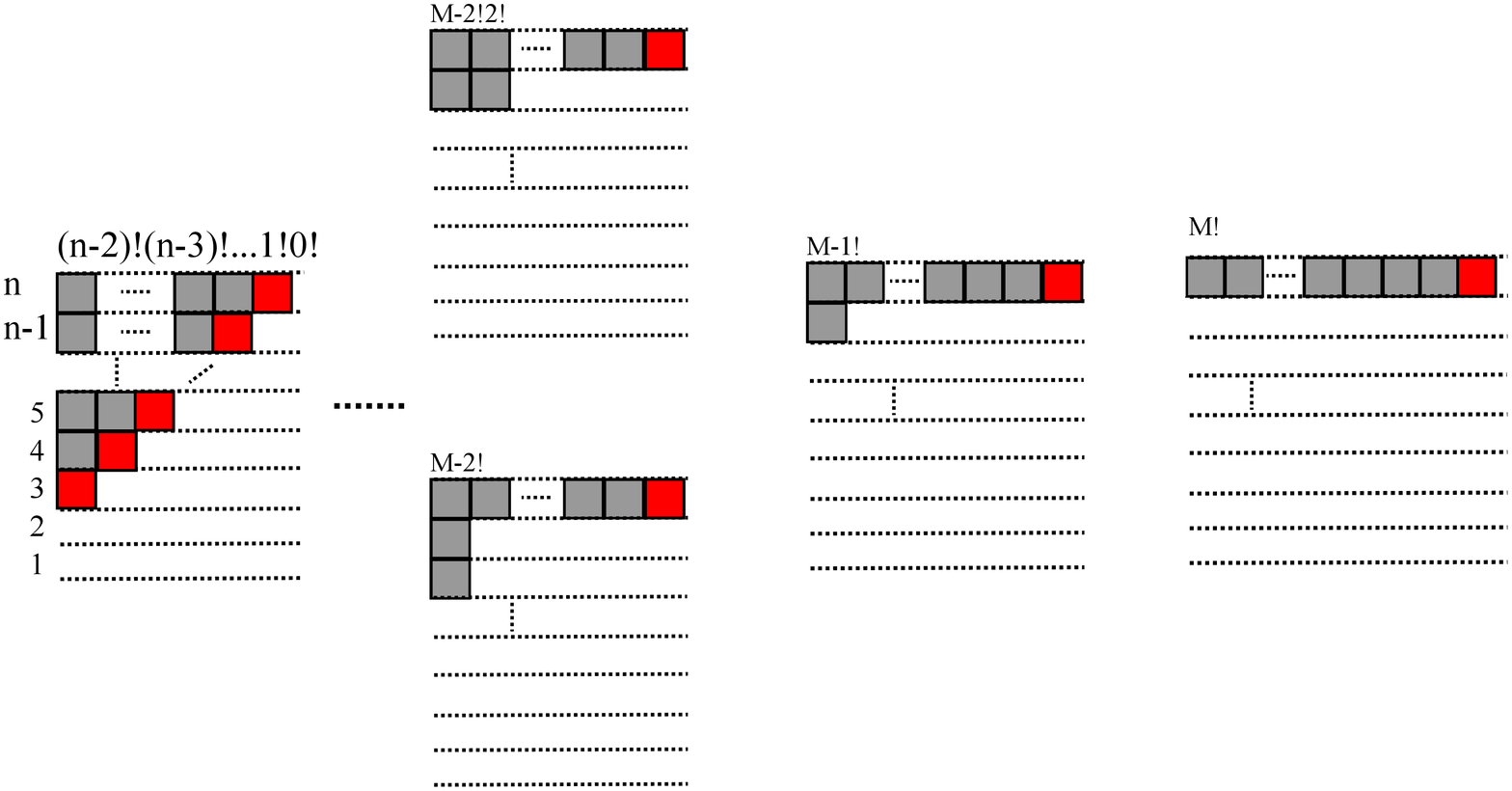}
  \caption{The tableaux of the non-zero $a$'s for the prepared form in n-point amplitude.}
  \label{fig:NPoinAllTab}
\end{figure}

\paragraph{Summary of the coefficient generating method}
Now let us summarize our method that computes the coefficients $a_{r_1,\cdots, r_n}$ in the differential operator $\mathbb D$ for a prepared CHY-type expression for $n$ points with $h_n = \sigma_i$. 

In Figure.\ref{fig:NPoinAllTab} we show the flowchart indicating the order in which the coefficients are solved. As discussed above, among the non-vanishing coefficients, the lowest-rank ones are elementary whose tableaux are depicted as the leftmost one in Figure.\ref{fig:NPoinAllTab} and their analytic solutions are given in (\ref{eq:elementSol}). These elementary coefficients are our building blocks and from their corresponding tableaux we move towards the right end of the flowchart. The tableaux for the remaining coefficients are all painted in the preferred way and thus the solutions of these non-elementary coefficients following from the corresponding relation read (supposing that there are $k$ red tiles that sit in the rows $\{i_1,\cdots, i_k\}$ while the the number of the unpainted tiles in each row is $r_i$),
\begin{align}
& a_{r_1,\cdots, r_{(i_1-1)}, (r_{i_1}+1), r_{(i_1+1)}, \cdots r_{(i_k+1)}, (r_{i_k}+1), r_{(i_k+1)}, \cdots, r_n } \nonumber\\
= &\frac{-1}{\ell_{i_1\cdots i_k} R(a_{r_1,\cdots, r_{(i_1-1)}, (r_{i_1}+1), r_{(i_1+1)}, \cdots r_{(i_k+1)}, (r_{i_k}+1), r_{(i_k+1)}, \cdots, r_n  })} \nonumber\\
&\times \left[ \sum_{j_1< \cdots < j_k, ~\{j\}\neq \{i\}} l_{j_1\cdots j_k} R(a)\, a_{r_1,\cdots, r_{(j_1-1)}, (r_{j_1}+1), r_{(j_1+1)}, \cdots r_{(j_k+1)}, (r_{j_k}+1), r_{(j_k+1)}, \cdots, r_n  }\right]\,,
\end{align}
where $R(a)$ denotes the rank of the coefficient it multiplies.
According to the previous discussion, in this case we always have $R(a) <R(a_{\cdots (r_{i_1}+1) \cdots (r_{i_k}+1) \cdots }) $ for any combination of $\{j_1,\cdots, j_k\} $ since $a_{\cdots (r_{i_1}+1) \cdots (r_{i_k}+1) \cdots } $ is non-elementary.

The relations corresponding to the painted tableaux in Figure.\ref{fig:NPoinAllTab} are independent from each other, since every relation involves a new coefficient. These relations are obviously complete as well, since they are linear and fix all the coefficients uniquely.
%
%
%

\section{One-loop five-point SYM amplitude}
\label{sec:FivePoint}

In this and the next sections, we further illustrate the aforementioned method by discussing more examples, the one-loop MHV amplitude for five external particles in $\mathcal N=4$ SYM and the $n$-gon amplitude for an arbitrary number of particles. At first sight, the one-loop SYM integrand \cite{Geyer:2015bja} might appear considerably more complicated than the four-point case due to the existence of the Pfaffian. Fortunately for us, supersymmetries, which enter the story in the GSO projection, simplify the integrand/amplitude by a great deal and cast the integrand into a form that is no longer intimidating at all. We first demonstrate this process using the relations among the Jacobi theta functions studied in \cite{Broedel:2014vla} and the decomposition of the polarization vectors. In the end we arrive at an expression that is a sum of prepared forms and the evaluation of this expression becomes straightforward.

\subsection{Simplify the one-loop generalized CHY integrand}
The generalized CHY form for the one-loop 5-point integrand in SYM is first conjectured as the following by Geyer \textit{et al} in~\cite{Geyer:2015bja, Geyer:2015jch} and checked against the corresponding Q-cut expressions up to five points,
\begin{eqnarray}\label{eq:MasonI5}
\mathcal I_5^{l=1} =\oint {d\sigma_1\cdots d\sigma_{4} \over  f_1\cdots f_{4}} ~\mathcal{F}~PT_5 \prod_{i=1}^{4} {1\over \sigma_i} \,,
\end{eqnarray}
where $PT_5$ is the well-known Parke-Taylor factor for five points and its general expression reads
\begin{equation}
    PT_n=\sum_{\rho\in \mathcal{S}_n}{1\over \sigma_{\rho(1)} (\sigma_{\rho(1)}-\sigma_{\rho(2)})(\sigma_{\rho(2)}-\sigma_{\rho(3)})\cdots(\sigma_{\rho(n-1)}-\sigma_{\rho(n)})}\,,
\end{equation}
in which $\mathcal{S}_n$ is the cyclic permutation group of $n$ elements.
The shorthand notation $\mathcal F$ denotes the summation of the Pfaffians over the GSO sectors that have even spin structures, namely $\mathcal{F}=\sum_{\alpha}c_\alpha \mathcal{Z}_\alpha Pf(W_\alpha)\left|_{\tau\rightarrow\infty}\right.$ where $\tau$ is regarded simply as a parameter in the SYM computations, but can be interpreted as the moduli parameter of the corresponding one-loop worldsheet in string theory.\footnote{ {This limit is equivalent to the $q\rightarrow 0$ limit in~\cite{Geyer:2015bja}.}} $\mathcal Z_\alpha$ denotes the theory-specific partition functions for the GSO sectors and in the case of $\mathcal N=4$ super Yang-Mills reads
\begin{align}
\mathcal{Z}_\alpha={\theta_\alpha(0|\tau)^4\over \eta(\tau)^{12}}\,, ~~~~\alpha=2,3,4
\end{align}
where $\eta(\tau)$ is the Dedekind eta function and $\alpha$ labels the GSO sectors with even spin structures, that is, $(R+),(NS+),(NS-)$ respectively. The coefficients $c_\alpha$ take care of the GSO projection, and thus take the values $\{c_2, c_3, c_4\}=\{+1, -1, +1\}$. The matrix in the Pfaffian takes the following skew-symmetric form, 
\begin{align}\label{eq:pfDef}
W_\alpha=\left(
\begin{array}{cc}
  A& -C^{T}    \\
  C&  B   
\end{array}
\right),
\end{align}
where 
\begin{align}
A_{ij}=\left\{
\begin{array}{c}
 \epsilon_i\cdot \epsilon_j \mathcal{S}_\alpha(z_{ij}|\tau)  \\
  0 
\end{array}
\right.\,, \;
B_{ij}=\left\{
\begin{array}{c}
 k_i\cdot k_j \mathcal{S}_\alpha(z_{ij}|\tau)  \\
  0 
\end{array}
\right.\,,\;
C_{ij}=\left\{
\begin{array}{cc}
 \epsilon_i\cdot k_j \mathcal{S}_\alpha(z_{ij}|\tau) & i\neq j \\
  -\epsilon_i\cdot F_i(z_i) &  i=j  
\end{array}
\right.\,,
\end{align}
where the polarization vectors and the momenta of the external gluons are denoted as $\epsilon$'s and $k$'s (without loss of generality, we choose the helicities of $\epsilon_1$ and $\epsilon_2$ to be negative and the rest positive throughout this section). $\mathcal S_\alpha$ denotes the fermionic propagator in the RNS formalism of string theory, with $\alpha$ labeling the sector, whose explict form will not be used in this paper.
We have also re-parameterized the variables as 
$\sigma=e^{2\pi i(z-{\tau\over 2})}$ and the scattering equations in terms of the complex variables $z_i$ read
\begin{equation}
\label{eq:ScatEq}
f_i\equiv \left.{k_i\cdot F_i \over \sigma_i}\right|_{\tau\rightarrow\infty} \,, ~~~k_i\cdot F_i(z_i)=k_i\cdot \ell+{1\over 2\pi i}\sum_{j\neq i} k_i\cdot k_j \mathcal R_{ij}\,,
\end{equation}
where $\mathcal R_{ij}\equiv -\partial_{z_i}\mathcal G(z_{ij}|\tau)={\theta'_1(z_{ij}|\tau)\over \theta_1(z_{ij}|\tau)}$, with $\mathcal G$ denoting the bosonic propagator.

Now we begin to simplify the Pfaffian (\ref{eq:pfDef}), which by definition is a polynomial in the elements of the matrix (\ref{eq:pfDef}). The terms in this polynomial are all of the following form with coefficients containing the kinetic informations only,
\begin{align*}
c(\{\epsilon\}, \{k\})\mathcal{S}^\alpha_{i_1 i'_1 i_2 i'_2 \cdots i_m i'_m}C_{j_1j_1}\cdots C_{j_{5-m} j_{5-m}} \,,
\end{align*}
where $\mathcal S^\alpha_{i_1 i'_1 i_2 i'_2 \cdots i_m i'_m} = \mathcal S_\alpha(z_{i_1 i'_1}|\tau)\cdots \mathcal S_\alpha(z_{i_m i'_m}|\tau)$ and $(i'_1, \cdots, i'_m)$ is a permutation of $(i_1, \cdots, i_m)$. 
The diagonal terms $C_{ll}$ of the off-diagonal blocks in (\ref{eq:pfDef}) are the only ones that contribute factors other than $\mathcal S_\alpha$. The union of all the indices above is simply $\{1,2,3,4,5\}$ with each number appearing twice while the sets of indices $\{ i_1, \cdots , i_m\}$ and $\{ j_1 ,\cdots j_{5-m} \}$ have no overlap. 
The coefficients $c(\{\epsilon\}, \{k\})$ and the entries $C_{ll}$ are evidently universal for all GSO sectors {and} can be pulled out from the the summation over the spin structures; the only factors relevant for the GSO summation are the products of $\mathcal S_\alpha$'s. Such summations are worked out explicitly at one loop for the $\mathcal N=4$ partition functions in \cite{Broedel:2014vla}, exploiting the properties of the Jacobi theta functions. For $m\leq 3$ the weighted sum simply vanishes and thus the surviving terms in the five-point Pfaffian are those with $m=4$ or $m=5$, which sum up to the following simple results when $z_{i_1 i'_1}+\cdots z_{i_m i'_m}=0$ (this condition is satisfied trivially in our case), 
\begin{eqnarray} 
&& \sum_{\alpha}c_\alpha \mathcal Z_\alpha S^\alpha_{i_1 i'_1 \cdots i_4 i'_4} = (2\pi)^4 \,, \\
&& \sum_{\alpha}c_\alpha \mathcal Z_\alpha S^\alpha_{i_1 i'_1 \cdots i_5 i'_5} = (2\pi)^4 (\mathcal R_{i_1 i'_1}+ \cdots +\mathcal R_{i_5 i'_5})\,.
\end{eqnarray}

Using these relations, the sum of the Pfaffians corresponding to the MHV amplitude over the even spin structures can be directly written down as the following,
\begin{eqnarray}
\label{eq:PfaffSym}
\mathcal{F}=\sum_{\rho\{345\}} \langle 12\rangle ^2 [ \rho(3)\rho(4)]^2 \text{$\epsilon $}_{\rho(5)}\cdot \ell +\sum_{\rho\{2345\}}\frac{ \langle 12\rangle ^4 [\rho(2)\rho(3)] [ \rho(4)\rho(5)]^2}{2\pi i\langle 1\rho(2)\rangle\langle 1\rho(3)\rangle } \mathcal{R}_{\rho_b(2)\rho_b(3)},
\end{eqnarray}
where $\rho\{345\}$ and $\rho\{2345\}$ are the  permutation groups $S_3/S_2$\footnote{$S_2$ indicates the permutation of the first two indices. Following cases are similar.} and $S_4/S_2\times S_2$ respectively. This expression does not yet put all the external particles on the same footing. In the $\mathcal N=4$ SYM case, it is possible to make the cyclic symmetry manifest at the level of integrand, by projecting the polarization vectors $\epsilon^\mu_i$ onto the external momenta $k^\mu_i$'s and eliminating the loop momentum $\ell$ using the scattering equations (\ref{eq:ScatEq}). The decomposition of the polarization vector reads,
\begin{align}
\epsilon_i^\mu={\epsilon(\epsilon_i,k_2,k_3,k_4)\over \epsilon(k_1,k_2,k_3,k_4)}k_1^\mu+{\epsilon(k_1,\epsilon_i,k_3,k_4)\over \epsilon(k_1,k_2,k_3,k_4)}k_2^\mu+{\epsilon(k_1,k_2,\epsilon_i,k_4)\over \epsilon(k_1,k_2,k_3,k_4)}k_3^\mu+{\epsilon(k_1,k_2,k_3,\epsilon_i)\over \epsilon(k_1,k_2,k_3,k_4)}k_4^\mu\,,
\end{align}
where $\epsilon (v_1, v_2, v_3, v_4) = \epsilon_{\rho_1 \rho_2 \rho_3 \rho_4} v^{\rho_1}_1 v^{\rho_2}_2 v^{\rho_3}_3 v^{\rho_4}_4$ and $\epsilon_{\rho_1\rho_2\rho_3\rho_4}$ denotes the Levi-Civita tensor and $v^{\rho_i}_i$ denotes the $\rho_i$-th component of the vector $v_j$.
Now all the $\epsilon_i\cdot \ell$ terms can be rewritten in terms of $\mathcal{R}_{i,j}$'s using the scattering equations, leading to the following form of (\ref{eq:PfaffSym}) that treats all the external lines democratically,
\begin{eqnarray}
\label{eq:PfaffSym2}
\mathcal{F}=\sum_{i\neq j}{\gamma_{[ij]}\over 2\pi i} \mathcal{R}_{ij},
\end{eqnarray}
where $\gamma_{[ij]}$ is defined in \cite{Carrasco:2011mn} as 
$$\gamma_{[\rho(1)\rho(2)]}=\langle 12\rangle^4 {[\rho(1)\rho(2)]^2[\rho(3)\rho(4)][\rho(4)\rho(5)][\rho(3)\rho(5)]\over 4\epsilon(\rho(1),\rho(2),\rho(3),\rho(4))},$$ with $\rho$ being a permutation element in $S_5/(S_2\times S_3)$.

Substituting (\ref{eq:PfaffSym2}) into (\ref{eq:MasonI5}) and taking the limit $\tau\rightarrow0$, the one-loop MHV amplitude for five points reads
\begin{eqnarray}
\label{eq:GCHYSimple}
\mathcal I_5^{l=1} =\oint {d\sigma_1\wedge\cdots \wedge d\sigma_{4} \over  f_1\cdots f_{4}} ~PT_5 \left(~\sum_{i\neq j} {\gamma_{[ij]}\over 2} {\sigma_i+\sigma_j\over \sigma_i-\sigma_j}\right) \prod_{r=1}^{4} {1\over \sigma_r} \,.
\end{eqnarray}

\subsection{Transformations to the prepared CHY integrand}
So far we have reduced the Pfaffian in the SYM integrand to a simple expression. Substituting the explicit expressions for the Parke-Taylor factor and applying the linear transformations that turn the original scattering equations to polynomials, we write down the following expressions for the five-point integrand,
\begin{eqnarray}
\label{eq:GCHYSimple2}
\mathcal I_5^{l=1} =\oint {d\sigma_1\wedge\cdots \wedge d\sigma_{4} \over  h_1\cdots h_{4}} \prod^5_{s-r\geqslant2} (\sigma_r-\sigma_s)\left(\sum_{k=1}^{5} {\sigma_{k+1}-\sigma_k\over \sigma_{k+1}}\right) \left(~\sum_{i\neq j} {\gamma_{[ij]}\over 2} {\sigma_i+\sigma_j\over \sigma_i-\sigma_j}\right)\,.
\end{eqnarray}
The factor 
$$\left(\sum_{k=1}^{5} {\sigma_{k+1}-\sigma_k\over \sigma_{k+1}}\right) \left(~\sum_{i\neq j} {\gamma_{[ij]}\over 2} {\sigma_i+\sigma_j\over \sigma_i-\sigma_j}\right)$$ 
is holomorphically equivalent\footnote{{ If two meromorphic forms differ from each other by  a holomorphic form, they are  holomorphically equivalent}} to 
$$
\sum_{i=1}^5\beta_{i,i+1,i+2,i+3,i+4}{\sigma_{i}-\sigma_{i-1}\over 2\sigma_{i}}+\sum_{i,i+1}\gamma_{[i,i+1]}{\sigma_i\over \sigma_i-\sigma_{i+1}}\left(\sum_{j\neq i+1} {\sigma_j-\sigma_{j-1}\over \sigma_j}\right),
$$
where $\beta_{i_1 i_2 i_3 i_4 i_5}=\sum\limits_{r<s}\gamma_{[i_r i_s]}$.
The former term above is already of the prepared form (\ref{eq:prepared}) up to the global residue theorem (which only introduces an overall minus sign to the result), while the latter, although not yet of the form we want, can be easily massaged into one, using the following cross ratio identities derived in \cite{Cardona:2016gon},
\begin{eqnarray*}
{s_{ii+1}\over \sigma_i-\sigma_{i+1}}={-1\over (\sigma_{i+1}-\sigma_{i+4})}\left({s_{ii+2}(\sigma_{i+2}-\sigma_{i+4})\over\sigma_{i}-\sigma_{i+2}}+{s_{ii+3}(\sigma_{i+3}-\sigma_{i+4})\over\sigma_{i}-\sigma_{i+3}}+{l_{i}(\sigma_{i}-\sigma_{i+4})\over\sigma_{i}}\right),
\end{eqnarray*}
where $i\in[1,5]$. The denominator on the right-hand side contains the factors $(\sigma_i-\sigma_j)$ of non-adjacent pairs $(ij)$ only and these factors are canceled out by the Vandermonde determinant that was introduced by the linear transformations of the scattering equations, leaving the monomial $\sigma_i$ in the denominator, which is just the case of the prepared form up to the global residue theorem.

Collecting all the prepared forms, the integrand (\ref{eq:GCHYSimple2}) is rearranged as follows,
\begin{eqnarray}
\label{eq:GCHYSimple2}
\mathcal I_5^{l=1} &=&\oint {d\sigma_1\wedge\cdots \wedge d\sigma_{4} \over  h_1\cdots h_{4}} \prod^5_{s-r\geqslant2} (\sigma_r-\sigma_s)\left(\sum_{i=1}^5\mathcal{P}_i+\sum_{j\neq i+1}^5\mathcal{B}_{[i,i+1],j}\right),
\end{eqnarray}
where 
\begin{align}
\mathcal{P}_i=\beta_{i,i+1,i+2,i+3,i+4}{\sigma_{i}-\sigma_{i-1}\over 2\sigma_{i}} \,,
\end{align}
and
\begin{align}
\mathcal{B}_{[i,i+1],j}=\left(\sum_{r=2}^3{s_{ii+r}\sigma_i(\sigma_{i+r}-\sigma_{i+4})\over s_{ii+1}(\sigma_{i+1}-\sigma_{i+4})(\sigma_{i+r}-\sigma_{i})}-{l_{i}(\sigma_{i}-\sigma_{i+4})\over s_{ii+1}(\sigma_{i+1}-\sigma_{i+4})}\right)\left({\sigma_j-\sigma_{j-1}\over \sigma_j}\right) \,.
\end{align}

\subsection{Evaluating  the prepared form of the five point integrand }
In order to evaluate (\ref{eq:GCHYSimple2}), we first construct the differential operators that capture the residues corresponding to the prepared expressions obtained previously.
Without loss of generality, we take the following prepared form with $h_5=\sigma_1$ as an example to demonstrate the detailed evaluation (all the other terms can be computed the same way),
\begin{eqnarray}\label{OneLoopFivePointGen}
 \mathcal I_5^{l=1}= \oint\limits_{h_1=\cdots=h_{4}=h_{5}=0} {d\sigma_1\wedge\cdots\wedge d\sigma_{4} \wedge d\sigma_5\over  h_1\cdots h_{4}h_5}  {\mathcal{H}(\sigma)}\,,
\end{eqnarray}
where $\mathcal{H}(\sigma)$ is a holomorphic function of the $\sigma_i$'s. The differential operator for this residue is of order 6 and has 210 coefficients $a_{r_1, r_2, r_3, r_4, r_5}$ to be determined.

We first get rid of those coefficients that are obviously zero. As discussed in the previous sections, these coefficients are $a_{r_1,\cdots, r_5}$ with $r_1\geq 1$ and those with ranks lower than $3! 2! 1! =12$, namely, $a_{0 , 1,1,1,3}$, $a_{0,1,1,2,2}$, and $a_{0, 0 ,2,2,2}$ (including permutations of the last four indices).
The total number of the vanishing coefficients is 
\begin{align}
C_8^3+C_7^3+C_6^3+C_5^3+C_4^3+C_3^3+C_4^1+C_4^2+C_4^1=140 \,.
\end{align}

\begin{figure}[htbp]
  \centering
  \includegraphics[width=0.75\textwidth]{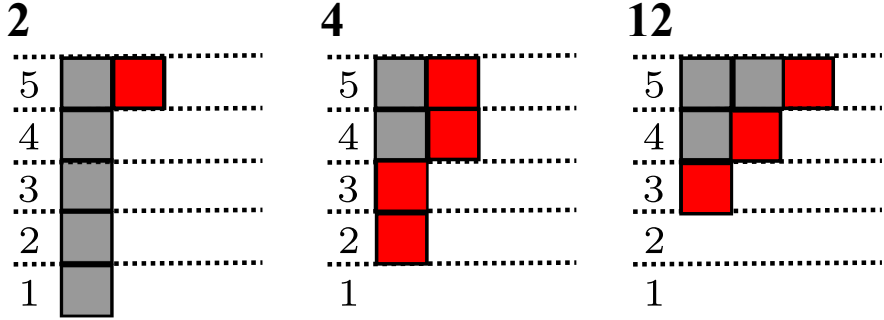}
  \caption{ The  tableaux for the elementary coefficients: $a_{2,1,1,1,1}, a_{2,2,1,1,0}, a_{3,2,1,0,0}$}\label{fig:FivePointS00}
\end{figure}

It is easy to observe that the elementary coefficients in this case are $a_{1,1,1,1,2}$, $a_{0,1,1,2,2}$ and $a_{0,0,1,2,3}$ (including index permutations) and their corresponding tableaux are depicted in Figure.\ref{fig:FivePointS00}. Among them, only $a_{0,0,1,2,3}$ (including the permutations of the last four indices) is non-vanishing. Therefore the intersection number constraint $\mathbb D\mathcal J=24$ yields
\begin{align}
12\sum_\rho (-1)^{\text{sgn}(\rho)} a_{0,\rho(0),\rho(1),\rho(2),\rho(3)} \ell_{v^{(\rho)}_3} \ell_{v^{(\rho)}_{2,3}} \ell_{v^{(\rho)}_{1,2,3}}\ell_{2,3,4,5}=24,
\end{align}
where $\rho$ runs over all permutations of $\{0,1,2,3\}$ and $v^{(\rho)}$ has the same meaning as in (\ref{eq:IntNPer}). For this particular case, these non-vanishing elementary coefficients read,
\begin{equation}
a_{0,\rho(0),\rho(1),\rho(2),\rho(3)}={ (-1)^{sgn(\rho)} \over 12\ell_{v^{(\rho)}_3} \ell_{v^{(\rho)}_{2,3}} \ell_{v^{(\rho)}_{1,2,3}}\ell_{2,3,4,5}}.
\end{equation}

\begin{figure}[htbp]
  \centering
  \includegraphics[width=0.95\textwidth]{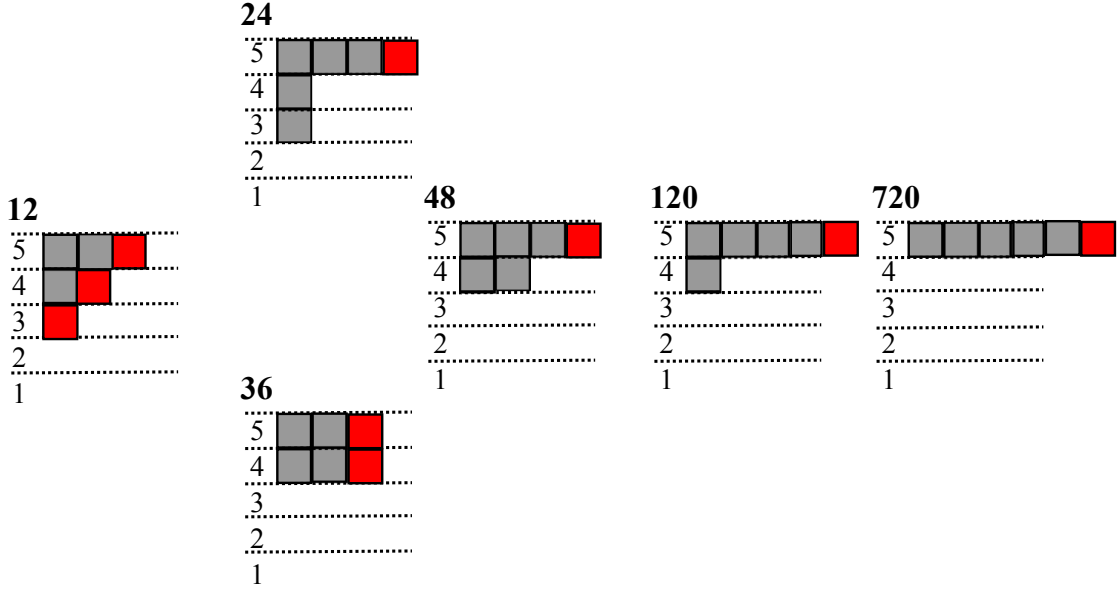}
  \caption{The tableaux of the non-zero $a$'s for the prepared form in five point amplitude}\label{fig:FivePointS1}
\end{figure}

Now we are only left with the non-elementary coefficients whose ranks are all higher than $12$, that is, $a_{0,0,1,1,4},~a_{0,0,0,2,2},~a_{0,0,0,2,4},~a_{0,0,0,1,5}$ and $a_{0,0,0,0,6}$. In Figure.\ref{fig:FivePointS1} their corresponding tableaux are shown and arranged in such an order that the preferred coloring relates each tableau with only the ones to the left of it.
The coefficients corresponding to the two tableaux in the second column in Figure.\ref{fig:FivePointS1}, namely $a_{0,0,1,1,4}$ and $a_{0,0,0,3,3}$, can be written in terms of the non-vanishing elementary one alone and the analytic solutions of them read,
\begin{align}
 a_{0,0,1,1,3+1} &= \frac{-1}{\ell_{5} 4!}\left[ \sum_{1\leqslant j_1\leqslant 5, ~\{j_1\}\neq \{5\}} l_{j_1} R(a) a_{r_{1}\cdots  (r_{j_1}+1)   \cdots r_5 }\right]\,,\nonumber\\
a_{0,0,0,2+1,2+1}&={-1\over 3!3!\ell_{4,5}}\left[ ~~~\sum_\updown{1\leqslant j_1<j_2\leqslant 5}{ ~\{j_1,j_2\}\neq \{4,5\}} ~~~l_{j_1,j_2} R(a) a_{r_{1}\cdots (r_{j_1}+1) \cdots (r_{j_2}+1)  \cdots r_5 }\right]\,.\nonumber
\end{align}
in which $\vec{r}=(0,0,1,1,3) ~\text{and}~ (0,0,0,2,2)$ respectively.
The coefficients associated with the other three tableaux are worked out in the same fashion and take the following forms,
\begin{align}
 a_{r_1,r_2,r_3,r_4,r_5+1} &=\frac{-1}{\ell_{5} R(a_{r_1,r_2,r_3,r_4,r_5+1})}\left[ \sum_{1\leqslant j_1\leqslant 5, ~\{j_1\}\neq \{5\}} l_{j_1} R(a) a_{r_{1}\cdots  (r_{j_1}+1)   \cdots r_5 }\right]\,,\nonumber
 \end{align}
 where $\vec{r}=(0,0,0,2,3),(0,0,0,1,4),(0,0,0,0,5)$ respectively.

With all the coefficients fixed for the case of (\ref{OneLoopFivePointGen}), it is now straightforward to evaluate the $\mathcal P_1$ and $\mathcal B_{[1,2],1}$ terms in (\ref{eq:GCHYSimple2}), by acting with the differential operator $\mathbb D$. Explicitly, the $\mathcal P_1$ term becomes
\begin{eqnarray}
\mathcal I_5^{\mathcal{P}_1} &=&{-\beta_{1,2,3,4,5}\over 2}\mathbb{D}\left.\frac{\sigma _3\sigma _4\sigma_5 \left(\sigma _2-\sigma _4\right) \left(\sigma _2-\sigma _5\right) \left(\sigma _3-\sigma _5\right)}{\sigma_5-1}\right|_{\sigma\rightarrow 0}={\beta_{1,2,3,4,5}\over 2}\mathbb{D}\left(\sigma _3\sigma^2_4\sigma^3_5 \right)\left|_{\sigma\rightarrow 0}\right.\nonumber\\
&=&{3!\beta_{1,2,3,4,5}}a_{0,0,1,2,3}={\beta_{1,2,3,4,5}\over 2\ell_{5}\ell_{4,5}\ell_{3,4,5}\ell_{2,3,4,5}}.
\end{eqnarray}
The integrand term containing $\mathcal{B}_{[1,2],1}$ is holomorphically equivalent to
\begin{eqnarray}
\label{eq:GCHYB121}
\mathcal I_5^{\mathcal{B}_{[1,2],1}} &=&\left(-{\gamma_{[12]}\ell_{1}\over s_{12}}\right)\oint {d\sigma_1\wedge\cdots \wedge d\sigma_{4} \over  h_1\cdots h_{4}\sigma_1}\sigma _3\sigma _4\sigma_5^2 \left(\sigma _2-\sigma _4\right) \left(\sigma _3-\sigma _5\right)\nonumber\\
&=&\left({\gamma_{[12]}\ell_{1}\over s_{12}}\right) \mathbb{D}\left({\sigma _3\sigma^2_4\sigma_5^3 \over \sigma_5-1}\right)=\left({\gamma_{[12]}\ell_{1}\over s_{12}}\right)3!2!a_{0,0,1,2,3}={-\gamma_{[12]}\over s_{12}\ell_{5}\ell_{4,5}\ell_{3,4,5}}
\end{eqnarray}

All other terms are similar.  The final results for all the terms are shown in Table~\ref{tab:FR1L5P} and \ref{tab:FR1L5B}, where the correspondence between each term and a particular forward-limit channel in the Q-cut analysis~\cite{Baadsgaard:2015twa} is displayed.
\begin{table}[htp]
\begin{center}
\caption{Final results for the pentagon diagrams. The first line lists the terms while the second line their analytic expressions respectively. The last line shows the forward-limit channels corresponding to each term, where the cut lines in the q-cut analysis are depicted in Fig. \ref{fig:FivePointPen}. For a given cut line, its momentum is chosen to be $\ell$.}
\begin{tabular}{|c|c|c|c|c|}\hline
$\mathcal{P}_1$&$\mathcal{P}_2$&$\mathcal{P}_3$&$\mathcal{P}_4$&$\mathcal{P}_5$\\ \hline
$\frac{\beta _{1,2,3,4,5}}{\ell _5 \ell _{4,5} \ell _{3,4,5} \ell _{2,3,4,5}}$&$\frac{\beta _{2,3,4,5,1}}{\ell _1 \ell _{1,5} \ell _{1,4,5} \ell _{1,3,4,5}}$&$\frac{\beta _{3,4,5,1,2}}{\ell _2 \ell _{1,2} \ell _{1,2,5} \ell _{1,2,4,5}}$&$\frac{\beta _{4,5,1,2,3}}{\ell _3 \ell _{2,3} \ell _{1,2,3} \ell _{1,2,3,5}}$&$\frac{\beta _{5,1,2,3,4}}{\ell _4 \ell _{3,4} \ell _{2,3,4} \ell _{1,2,3,4}}$\\ \hline
$E1$&$E2$&$E3$&$E4$&$E5$\\ \hline
\end{tabular}
\label{tab:FR1L5P}
\end{center}
\end{table}%
 \begin{figure}[htbp]
  \centering
  \includegraphics[width=0.35\textwidth]{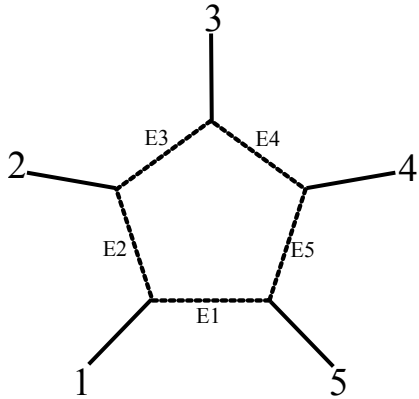}
  \caption{Pentagon Diagram.}\label{fig:FivePointPen}
\end{figure}
 
 \begin{table}[htp]
\begin{center}
\caption{Final results for the box diagrams, $i\in [1,5]$. Like in Table.\ref{tab:FR1L5P}, the first and the second lines show the terms and their respective final expressions, while in the last line their corresponding forward-limit channels are given and the cut lines are depicted in Figure.\ref{fig:FivePointBox}. Again, for a given cut line in the q-cut analysis, its momentum is chosen to be $\ell$.}
\begin{tabular}{|c|c|c|c|}\hline
$\mathcal{B}_{[i,i+1],i}$&$\mathcal{B}_{[i,i+1],i+2}$&$\mathcal{B}_{[i,i+1],i+3}$&$\mathcal{B}_{[i,i+1],i+4}$\\ \hline
$\frac{-\gamma_{i,i+1}/s_{i,i+1}}{\ell _{i+4}  \ell _{i+3,i+4} \ell _{i+2,i+3,i+4}}$&$\frac{-\gamma_{i,i+1}/s_{i,i+1}}{\ell _{i+3}  \ell _{i+2,i+3} \ell _{i,i+1,i+2,i+3}}$&$\frac{-\gamma_{i,i+1}/s_{i,i+1}}{\ell _{i+2} \ell _{i,i+1,i+2} \ell _{i,i+1,i+2,i+4}}$&$\frac{-\gamma_{i,i+1}/s_{i,i+1}}{\ell _{i,i+1} \ell _{i,i+1,i+4} \ell _{i,i+1,i+3,i+4}}$\\ \hline
$E1$&$E3$&$E4$&$E5$\\ \hline
\end{tabular}
\label{tab:FR1L5B}
\end{center}
\end{table}%
 \begin{figure}[htbp]
  \centering
  \includegraphics[width=0.35\textwidth]{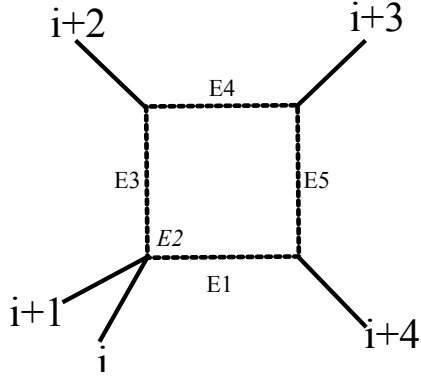}
  \caption{Box Diagram.}\label{fig:FivePointBox}
\end{figure}

This concludes the computation of the residues at finite poles and at infinity  (the computation of the residue at infinity is transformed to evaluating a prepared form with $h_n = \sigma_n$ in the same fashion as in~\cite{Chen:2016fgi}).

\subsection{Analysis for the spurious poles}
\label{sec:Spurious5ptSYM}
In this subsection we investigate {the} spurious poles that arise from the extra solutions to the polynomial scattering equations. Such contribution should vanish in consistent supersymmetric theories and we will show this is indeed true in this particular case. It is easy to observe that the extra solutions all coincide at the position $\sigma_1=\sigma_2=\cdots=\sigma_{n-1}=1$. Naturally, we prefer to deal with poles at the origin and shift the spurious pole there by the parameter transformation $\sigma_i \rightarrow \sigma_i+1$. Therefore the residue at the spurious pole is associated with the meromorphic form,
\begin{align}
\rho_{spurious} = \frac{d\sigma_1\wedge d\sigma_2 \wedge d\sigma_3 \wedge d\sigma_4}{h_1(\sigma_j+1) h_2 (\sigma_j+1) h_3(\sigma_j+1) h_4 (\sigma_j+1)} \mathcal H_5 (\sigma_j+1)\,,
\end{align}
where the factor $\mathcal H_5$ is read off from (\ref{eq:GCHYSimple2}) and 
\begin{align}\label{H5shifted}
\mathcal H_5(\sigma_j+1)=\sum_{r<s}\beta_{[rs]}\sum_{\rho\in S_5}{\prod_{i<j}^5(\sigma_i-\sigma_j)\over (\sigma_{\rho(1)}+1) (\sigma_{\rho(1)}-\sigma_{\rho(2)})\cdots(\sigma_{\rho(4)}-\sigma_{\rho(5)})}  {\sigma_r+\sigma_s+2\over\sigma_r-\sigma_s}\,.
\end{align}
The product $\prod_{i+1<j} (\sigma_i-\sigma_j)$ contains all non-adjacent pairs of $(\sigma_i-\sigma_j)$ with the gauge choice being implicitly $\sigma_5=1$.
The shifted polynomial scattering equations read the following,
\begin{eqnarray}
h_1 (\sigma_j+1) &=& \sum_{i=1}^5 \ell_i+ \sum_{i=1}^4 \ell_i \sigma_i \,, \nonumber\\
h_2 (\sigma_j+1) &=& -\sum_{i_1<i_2}^{5} \ell_{i_1, i_2}-\sum_{i_1=1}^{4}\sum_{i_2\neq i_1}^5\ell_{i_1,i_2}\sigma_{i_1}-\sum_{i_1<i_2}^{4} \ell_{i_1,i_2}\sigma_{i_1} \sigma_{i_2} \,, \nonumber\\
h_3 (\sigma_j+1) &=& \sum_{i_1<i_2<i_3}^{5} \ell_{i_1, i_2, i_3}+\sum_{i_1=1}^{4}\sum_{i_2<i_3\neq i_1}^5\ell_{i_1,i_2,i_3}\sigma_{i_1}+\sum_{i_1<i_2}^{4}\sum_{i_3\neq i_1, i_2}^5\ell_{i_1,i_2,i_3}\sigma_{i_1}\sigma_{i_2}+\sum_{i_1<i_2<i_3}^{4} \ell_{i_1,i_2,i_3}\sigma_{i_1} \sigma_{i_2} \sigma_{i_3}\,, \nonumber\\
h_4 (\sigma_j+1) &=& -\sum_{i_1<\cdots < i_4}^5 \ell_{i_1, i_2, i_3, i_4} -\sum_{i_1=1}^4 \sum_{i_2<i_3<i_4\neq i_1}^5 \ell_{i_1, i_2, i_3, i_4} \sigma_{i_1} - \sum_{i_1<i_2}^4 \sum_{i_3<i_4\neq i_1, i_2}^5 \ell_{i_1,i_2, i_3, i_4} \sigma_{i_1} \sigma_{i_2}  \nonumber\\
 &&-\sum_{i_1< i_2< i_3}^4 \sum_{i_4\neq i_1,i_2 i_3}^5 \ell_{i_1,i_2,i_3,i_4} \sigma_{i_1}\sigma_{i_2} \sigma_{i_3} -\sum_{i_1<\cdots i_4}^4 \ell_{i_1,i_2, i_3, i_4} \sigma_{i_1} \sigma_{i_2} \sigma_{i_3} \sigma_{i_4}\,.\nonumber
\end{eqnarray} 
Note that the leading order terms above do \textit{not} give rise to a zero-dimensional intersection and we have to apply transformations on these polynomials such that the divisors generated by them coincide at an isolated point.
The explicit expressions for such transformations will not be needed and we simply display the new polynomials here,
\begin{eqnarray}
\hat h_1 &=&  \sum_{i=1}^4 \ell_i \sigma_i \,, \nonumber\\
\hat h_2 &=&  -\sum_{i_1<i_2}^4 \ell_{i_1, i_2} \sigma_{i_1} \sigma_{i_2} \, , \nonumber \\
\hat h_3 &=& \sum_{i_1<i_2}^4 \sum_{i_3\neq i_1, i_2}^5 \ell_{i_1, i_2, i_3} \sigma_{i_1} \sigma_{i_2} + \sum_{i_1<i_2< i_3}^4 \ell_{i_1, i_2, i_3} \sigma_{i_1}\sigma_{i_2}\sigma_{i_3}\, ,\nonumber\\ 
\hat h_4 &=&  -\left( \sum_{i_1<i_2< i_3}^4 \sum_{i_4\neq i_1,i_2, i_3}^5 \ell_{i_1, i_2, i_3, i_4} \sigma_{i_1}\sigma_{i_2} \sigma_{i_3} + \sum_{i_1<\cdots i_4}^4 \ell_{i_1, i_2 , i_3, i_4} \sigma_{i_1}\sigma_{i_2} \sigma_{i_3}\sigma_{i_4}\right)\,.\nonumber
\end{eqnarray}
According to the transformation law, we also need to take care of the determinant of the transformations, which is simply $1$ here. 

In principle we need to construct a different differential operator following Section 4.4 of~\cite{Chen:2016fgi}; and to this end the hatted polynomials are homogenized in a way such that the higher-order terms are \textit{lowered} to the same degree as the leading order ones, leading to the order of this operator being $4$. Luckily the actual form of the operator is not necessary in this case, since (\ref{H5shifted}) is holomorphic at the shifted origin and the lowest term in its numerator is of degree 5. The action of the fourth-order differential operator on (\ref{H5shifted}) must vanish when evaluated at the shifted origin.


Since the spurious poles do not contribute, the total residue obtained previously is the final integrand for the five-point SYM amplitude at one loop which simply reads,
\begin{align}
\mathcal I^{l=1}_5 = \sum_{i=1}^5 \mathcal I^{\mathcal P_i}_5 +\sum_{j\neq i+1}^5 \mathcal I^{\mathcal B_{[i,i+1], j}}_5 \,,
\end{align}
where the expressions for $\mathcal I^{\mathcal P_i}_5$ and $\mathcal I^{\mathcal B_{[i,i+1],j}}_5$ are given in Table \ref{tab:FR1L5P} and Table \ref{tab:FR1L5B}. Just like its four-point counterpart studied in~\cite{Chen:2016fgi}, our result presents a clear one-to-one correspondence with the forward-limit channels in the Q-cut analysis. We expect this nice property continues to hold for higher points.   

The method we have illustrated so far generalizes naturally to the one-loop integrands for any number of external particles. For higher-point cases, we expect the integrands can be decomposed into several residues that are associated with the prepared form, or similar expressions with slightly more complicated $h_n$'s.

Similar to the analysis in Section.\ref{sec:prescription}, when $h_n$ is a monomial, the corresponding local duality theorem leads to the vanishing of some coefficients. 

The local duality equations arising from the scattering equations are universal despite the choice of $h_n$, that is to say, the fact that the coefficients in the corresponding differential operator can be related to those with lower or equal ranks only holds for any number of points, and most of the relations among these coefficients remain the same. The construction of the differential operator always boils down to solving for the elementary coefficients. Since the number of the elementary coefficients is quite limited, these coefficients can be solved efficiently.

\section{One-loop n-gon amplitude: A direct evaluation}
\label{sec:n-gon}
In this section we discuss the evaluation of the generalized CHY form for  one-loop n-gon amplitude. The evaluation can be done directly since all n-gon amplitudes are already of the prepared form. The one-loop n-gon amplitude is conjectured to be \cite{Cachazo:2015aol}
\begin{eqnarray}\label{eq:NGON}
\mathcal I_n =\oint {d\sigma_1\cdots d\sigma_{n-1} \over  h_1\cdots h_{n-1}} ~PT_n \prod_{i<j}^n (\sigma_i-\sigma_j) \,,
\end{eqnarray}
where $PT_4$ is the Parke-Taylor factor.
The terms in (\ref{eq:NGON}), after homogenization, are of form 
$$\mathcal I_n=\sum_{\rho\in \mathcal{S}_n} \mathcal I_n^{\rho(1)},$$
where 
\begin{eqnarray}\label{eq:NGON2}
\mathcal I_n^{\rho(1)} =\oint {d\sigma_1\wedge\cdots\wedge d\sigma_{n} \over  \tilde h_1\cdots \tilde h_{n-1}\sigma_{\rho(1)}} ~\sigma_{\rho(n)}\prod_{i||j}^n (\sigma_{\rho(i)}-\sigma_{\rho(j)} )  \,,
\end{eqnarray}
where $i||j$ denotes pairs of neighboring indices in the cyclic order.
In terms of the differential operator representation, we have 
\begin{eqnarray}\label{eq:NGON2}
\mathcal I_n^{\rho(1)} &=&-\mathbb{D} ~{{-\sigma_{\rho(n)}\prod_{i||j}^n (\sigma_{\rho(i)}-\sigma_{\rho(j)} )}\over \sigma_{\rho(1)}-1 }\left|\right._{\sigma\rightarrow 0} \,,\nonumber\\
&=&\mathbb{D}{\sigma_{\rho(3)}^{1} \sigma_{\rho(4)}^{2} \cdots\sigma_{\rho(n)}^{n-2}\over 1-\sigma_{\rho(1)}} \nonumber\\
&=&a_{v^{(\rho^{-1})}_1,v^{(\rho^{-1})}_2,\cdots, v^{(\rho^{-1})}_n},
\end{eqnarray}
where $v^{(\rho^{-1})}$ means the result of performing $\rho^{-1}$ permutation on $v=\{0,0,1,2,3,\cdots, n-3,n-2\}$, and subscripts of $v$ indexes the components of $v^{(\rho^{-1})}$.These $a$'s have already been obtained in Eq. (\ref{eq:elementSol}). On the other hand, as we will see from later discussion, here the spurious pole does not contribute. Thus the sum over $\rho\in\mathcal S_n$ of (\ref{eq:NGON2}) is the final result of the general one-loop n-gon amplitudes.

\paragraph{The analysis for the spurious pole}
Here we show that in this case there is indeed no contribution from the spurious pole, which is located at $\sigma_1=\cdots=\sigma_{n-1}=1$. The parameter transformation $\sigma_j\rightarrow \sigma_j+1$ shifts the pole to the origin. The polynomial scattering equations are those appearing in (\ref{eq:polynomialScatEq}).
Therefore, what we want to compute here is the residue at the origin of the differential form,
\begin{eqnarray}
\rho_{\text{spurious}} = \frac{d\sigma_1 \wedge d\sigma_2 \cdots\wedge d\sigma_{n-1}}{h_1(\sigma_j+1) h_2(\sigma_j+1) \cdots h_{n-1}(\sigma_j+1)} \,\mathcal H_n(\sigma_j+1)\,
\end{eqnarray}
where
\begin{equation}
\label{eq:PolyScatSpurious}
h_m(\sigma_j+1)=(-)^{m+1}\sum_{r=0}^m\sum_{i_1<\cdots<i_r}^{n-1}~~~ \sum_{\updown{i_{r+1}<\cdots<i_m}{\neq i_1,\cdots, i_r}}^n \ell_{i_1,\cdots, i_m} \sigma_{i_1}\sigma_{i_2}\cdots \sigma_{i_r}.
\end{equation}
For instance, the first three of the scattering equations are explicitly
\begin{eqnarray}
&&h_1(\sigma_j+1)=\sum_{i=1}^n \ell_i+\sum_{i=1}^{n-1} \ell_i\sigma_i ,\nonumber\\
&&h_2(\sigma_j+1)=-\left(\sum_{i_1<i_2}^{n} \ell_{i_1, i_2}+\sum_{i_1=1}^{n-1}\sum_{i_2\neq i_1}^n\ell_{i_1,i_2}\sigma_{i_1}+\sum_{i_1<i_2}^{n-1} \ell_{i_1,i_2}\sigma_{i_1} \sigma_{i_2}\right) ,\nonumber\\
&&h_3(\sigma_j+1)=\sum_{i_1<i_2<i_3}^{n} \ell_{i_1, i_2, i_3}+\sum_{i_1=1}^{n-1}\sum_{i_2<i_3\neq i_1}^n\ell_{i_1,i_2,i_3}\sigma_{i_1}+\sum_{i_1<i_2}^{n-1}\sum_{i_3\neq i_1}^n\ell_{i_1,i_2,i_3}\sigma_{i_1}\sigma_{i_2}+\sum_{i_1<i_2<i_3}^{n-1} \ell_{i_1,i_2,i_3}\sigma_{i_1} \sigma_{i_2} \sigma_{i_3}. \nonumber
\end{eqnarray}
According to the momentum conservation and on-shell conditions,  we find the variety of the leading terms of them are not zero dimensional. For such cases, we need to perform a transformation law, then we have
\begin{eqnarray}
\rho_{\text{spurious}} = \frac{d\sigma_1 \wedge d\sigma_2 \cdots\wedge d\sigma_{n-1}}{\hat h_1 \hat h_2 \cdots \hat h_{n-1}} \,\mathcal H_n(\sigma_j+1)\,.
\end{eqnarray}
where the polynomials become 
\begin{eqnarray}
\label{eq:PolyScatSpuriousT}
\hat h_1&=&\sum_{i=1}^{n-1} \ell_i\sigma_i ,\nonumber\\
\hat h_2&=&-\sum_{i_1<i_2}^{n-1} \ell_{i_1,i_2}\sigma_{i_1} \sigma_{i_2} \nonumber\\
\hat h_m&=&(-)^{m+1}\sum_{r=m-1}^m\sum_{i_1<\cdots<i_r}^{n-1}~~~ \sum_{\updown{i_{r+1}<\cdots<i_m}{\neq i_1,\cdots, i_r}}^n \ell_{i_1,\cdots, i_m} \sigma_{i_1}\sigma_{i_2}\cdots \sigma_{i_r}, m\in [3,n-1].
\end{eqnarray}
The intersection number  of $\hat h_m$ at shifted origin is $2 (n-2)!$. This indicate that there are $2 (n-2)!$ solutions of $\sigma$  concentrating on the spurious point.  The residue on the shifted origin is well-defined only when $\mathcal H_n(\sigma_j+1)$ is holomorphic at the origin. For n-gon, 
$$\mathcal H_n(\sigma_j+1)=\sum_{\rho\in S_n}{\prod_{i<j}^n(\sigma_i-\sigma_j)\over (\sigma_{\rho(1)}+1) (\sigma_{\rho(1)}-\sigma_{\rho(2)})(\sigma_{\rho(2)}-\sigma_{\rho(3)})\cdots(\sigma_{\rho(n-1)}-\sigma_{\rho(n)})} \left|_{\sigma_n\rightarrow 0}\right.$$
It is easy to see that each factor $\sigma_{\rho(i)}-\sigma_{\rho(i+1)}$ in the dominator will be canceled by a factor in the numerator. The finial degree of $\sigma$ in the numerator is 
$${(n-2)(n-1)\over 2}.$$ 
According to our conjecture, the differential operator $\mathbb{D}$ for the residue at the shifted origin is 
$${(n-4)(n-1)\over 2}+2,$$ 
where $n\geqslant 4$.
Then it is easy to find that the spurious residue are all vanishing for n-gon with $n\geqslant 4$.


\section{Conclusion and outlook}
In this paper we use our previously proposed  differential operator 
 to compute  the residues on the solutions of scattering equations. 
We find that the coefficients in the  differential operator can 
be determined in a combinatoric way.
We present an analytical solution for the differential operator for 
 the prepared forms of the generalized CHY integrands for  $n$-point scattering amplitudes.
We then use said differential operator to evaluate the one-loop CHY integrand for five points in $\mathcal N=4$ SYM which is casted into the prepared forms, and the one-loop CHY integrands of the $n$-gon amplitudes which are naturally of the prepared forms for any number of external lines. In both examples, our final results are identical with the ones obtained  through the Q-cut analysis respectively.
 
Although all the examples studied in this paper can be massaged to the prepared form, more complicated denominators may turn up in the integrands when we investigate higher-point one-loop amplitudes in super Yang-Mills. An immediate followup is to generalize our method for differential operators corresponding to these new ingredients. Once such generalization is worked out, the evaluation of any one-loop integrand in SYM will become straightforward and economical.
   
At the moment of writing, the construction of generalized CHY forms in super Yang-Mills theory has been promoted to two loops for four external lines. Another natural direction to take is to study the combinatoric structures of the two-loop scattering equations and lift the coefficient-generating method to two loops, which in turn will provide a useful tool in constructing the generalized CHY form for any number of external lines at two-loop level.

Moreover, the method we propose is theory-independent and thus has the potential to evaluate any CHY-type expressions in pure Yang-Mills and Gravity theory. This, in turn, will make the generalized CHY forms an efficient method for computing amplitudes.
%

\acknowledgments
GC and TW thank E. Y. Yuan,  Y. Zhang, R.J. Huang, B. Feng  and H. Johansson for useful discussion and kind suggestions. 
GC  has been supported in parts by  NSF of China Grant under Contract 11405084, the Open Project Program of State Key Laboratory of Theoretical Physics, Institute of Theoretical Physics, Chinese Academy of Sciences (No.Y5KF171CJ1). 
Y-K.E.C acknowledges the European Union's Horizon 2020 research
 and innovation programme under the Marie Sk\'lodowska-Curie 
 grant agreement No 644121, and  the Priority Academic Program 
 Development for Jiangsu Higher Education Institutions (PAPD).

\bibliographystyle{JHEP}
\bibliography{ScatEq}
\end{document}